\documentclass[a4paper,fleqn,usenatbib]{mnras}

\usepackage[T1]{fontenc}
\usepackage{ae,aecompl}

\usepackage{graphicx}	
\usepackage{amsmath}	
\usepackage{amssymb}	
\usepackage{float}
\usepackage{natbib}
\usepackage{morefloats}
\usepackage{dcolumn}
\usepackage{amsmath}
\usepackage{latexsym}
\usepackage{longtable}
\usepackage{lipsum}
\usepackage{gensymb} 
\usepackage{adjustbox}
\usepackage{lscape}

\title[M12045]{Observational properties of a Type Ib Supernova MASTER OT J120451.50+265946.6 in NGC 4080}
\author[Mridweeka Singh et al.]{Mridweeka Singh$^{1,2}$\thanks{E-mail: mridweeka@aries.res.in, yashasvi04@gmail.com},
Kuntal Misra$^{1,3}$,
D.K.Sahu$^{4}$,
Raya Dastidar$^{1,5}$,
\newauthor
Anjasha Gangopadhyay$^{1,2}$,
Shubham Srivastav$^{6}$,
G. C. Anupama$^{4}$,
\newauthor
Subhash Bose$^{7}$,
Vladimir Lipunov$^{8,9}$,
N. K. Chakradhari$^{2}$,
Brajesh Kumar$^{4}$,
\newauthor
Brijesh Kumar$^{1}$,
S. B. Pandey$^{1}$,
Evgeny Gorbovskoy$^{9}$,
Pavel Balanutsa$^{9}$
\\
$^{1}$Aryabhatta Research Institute of observational sciencES, Manora Peak, Nainital 263 001, India\\
$^{2}$School of Studies in Physics and Astrophysics, Pandit Ravishankar Shukla University, Chattisgarh 492 010, India \\
$^{3}$Department of Physics, University of California, 1 Shields Ave, Davis, CA 95616-5270, USA \\
$^{4}$Indian Institute of Astrophysics, Koramangala, Bangalore 560 034, India\\
$^{5}$Department of Physics \& Astrophysics, University of Delhi, Delhi 110 007, India\\
$^{6}$Department of physics, Indian Institute of Technology, Powai, Mumbai 400076, India\\
$^{7}$Kavli Institute for Astronomy and Astrophysics, Peking University, 5 Yiheyuan Road, Haidian District, Beijing 100871, China\\
$^{8}$Lomonosov Moscow State University, Physics Department, 119991, Vorobievy hills, 1, Moscow, Russia \\
$^{9}$Lomonosov Moscow State University, SAI, 119234, Universitetsky pr., 13, Moscow, Russia \\
}

\date{Accepted XXX. Received YYY; in original form ZZZ}

\pubyear{2018}

\begin{document}
\label{firstpage}
\pagerange{\pageref{firstpage}--\pageref{lastpage}}
\maketitle

\begin{abstract}
MASTER OT J120451.50+265946.6 (M12045), discovered by the MASTER Global Robotic Net, is a Type Ib supernova (SN) that exploded in NGC 4080.  We present the {\it BVRI} photometric and spectroscopic observations upto $\sim$250 days since $B_{max}$. At the time of discovery the SN was a few weeks past maximum light and our observations capture the linearly declining light curve phase.  M12045 declines faster as compared to SNe 1999dn and 2009jf at comparable epochs.  Rigorous spectroscopic monitoring reveals that M12045 is a normal Type Ib SN.  The analysis of the nebular phase spectra indicates that $\sim$ 0.90 M$_\odot$ of O is ejected in the explosion.  The line ratio of [\ion{O}{I}] and [\ion{Ca}{II}] in the nebular phase supports a massive WR progenitor with main sequence mass of $\sim$ 20 M$_\odot$.
\end{abstract}




\section{Introduction}
\label{Introduction}

Type Ib supernovae (SNe) are a subclass of hydrogen deficient SNe with prominent He features in their early time spectra (for a review, see \citealt{1997ARA&A..35..309F}). On the other hand, Type Ic SNe are devoid of both H and He features. In the context of spectral properties, the presence of hydrogen in Type Ib/c SNe is still an open question. The absence of hydrogen in early spectra of Type Ib SNe does not imply a complete absence of hydrogen.  The absorption feature at $\sim$ 6200 \AA~in early spectra of these SNe is usually attributed to H$\alpha$ \citep{2002ApJ...566.1005B,2005ApJ...631L.125A,2008Natur.453..469S}.  Both \cite{2002ApJ...566.1005B} and \cite{2006A&A...450..305E} have emphasized the existence of hydrogen in Type Ib SNe in different studies conducted on Type Ib SNe.  Similar study on Type Ic SNe has also revealed the presence of Helium in some Type Ic SNe \citep{2006PASP..118..791B,2006A&A...450..305E}.   The signatures of H and He found in Type Ib/c SNe are due to a thin layer of these elements and their continuous stripping from the progenitor \citep{2011MNRAS.416.3138V}.

The progenitors of Type Ib/c SNe lose their hydrogen/helium envelope via strong stellar winds. There are two possible progenitor scenarios proposed for Type Ib/c SNe. One is a massive WR star ($>$ 20--25 M$_\odot$) which has lost its hydrogen envelope by transfer of mass to a companion star or by stellar winds \citep{1986ApJ...306L..77G}.  Another scenario is a low mass progenitor ($>$ 11 M$_\odot$) in a binary system \citep{1992ApJ...391..246P,1995PhR...256..173N,2009ARA&A..47...63S}. A direct identification of progenitor in the pre-SN images is usually a reliable way to distinguish between the different progenitor types  \citep{2009ARA&A..47...63S},  but previous attempts in this direction for Type Ib/c SNe have been unsuccessful \citep{2007MNRAS.381..835C,2009ARA&A..47...63S,2013MNRAS.436..774E}.   However,  in one of the most recent study \cite{2013ApJ...775L...7C} have reported the possible progenitor identification of iPTF13bvn within a 2$\sigma$ error radius consistent with a massive WR progenitor star.   The massive WR progenitor scenario is also supported by the stellar evolutionary models \citep{2013A&A...558L...1G}.  Based on observational evidences from early and nebular phase spectroscopy,   both a massive WR progenitor as well as an interacting binary progenitor is proposed by several authors \citep{2014A&A...565A.114F, 2016A&A...593A..68F, 2016ApJ...825L..22F,2014AJ....148...68B,2015MNRAS.446.2689E,2015A&A...579A..95K,2016MNRAS.461L.117E,2017MNRAS.466.3775H,2017MNRAS.469L..94H}.   \cite{2012A&A...544L..11Y} have derived the magnitudes of Type Ib/c progenitor stars at pre-SN stage using massive star evolutionary models and found them to be in the range of M$_{V}$ $\approx$ -2 $\sim$ -3 mag.  These are fainter than most of the WR stars in the nearby Universe.    Whereas helium star progenitors with low mass in binary systems are brighter in optical domain since they convert into He giant star \citep{2012A&A...544L..11Y}.  A similar conclusion is also given by \cite{2015MNRAS.446.2689E} for binary progenitors. In the case of type Ic SNe there is a first possible identification of a progenitor reported for SN 2017ein \citep{2018ApJ...860...90V}.  

Some of the Type Ib/c explosions are known to be aspheric. A reasonable degree of polarization and polarization angle, which depends upon the relative size of asymmetry and the sky orientation \citep{1982ApJ...263..902S,1984MNRAS.210..829M,1991A&A...246..481H},  is reported for these objects. The observed higher degree of polarization in the spectropolarimetric studies during early phase also confirms the asphericity  \citep{2003ApJ...592..457W,2006Natur.440..505L}.

In this paper we present the photometry and spectroscopy of MASTER OT J120451.50+265946.6  (hereafter M12045) upto $\sim$250 days since $B_{max}$ followed by a detailed analysis and interpretation of the SN characteristics.  

\section{Data Acquisition and Reduction}\label{sec:Data Acquisition and Reduction} 
MASTER Global Robotic Net consists of eight twin robotic telescopes \citep{2010AdAst2010E..30L}, that work in alert, inspection and survey mode of observations in every night with a clear sky. 
The main MASTER feature is own real-time auto-detection system \citep{2010AdAst2010E..30L}, that automatically reduces four square degrees in 1-2 minutes after CCD readout.   To discover new optical sources (transients) MASTER usually make unfiltered automatic survey in W=0.2B+0.8R, calibrated by USNO-B1 R2,B2/R1,B1 thousands field stars. 
During such survey MASTER-Tunka auto-detection system discovered optical transient (OT) source at RA = 12h04m51.50s and Dec = 26d59m46.6s on 2014-10-28.87454 UT with 13.9m unfiltered magnitude (magnitude limit = 18.1 mag). This OT was located 4$\arcsec$ West and 13$\arcsec$ North of the center of NGC 4080 galaxy. MASTER has reference image without OT on 2011-04-21.67477 UT with 20.2m unfiltered magnitude limit \citep{2010AdAst2010E..30L}.  The MASTER photometry of the OT is listed in Table \ref{tab:photometric_observational_log}.

Our multiband observing campaign of M12045 started 17 days after the MASTER discovery and continued upto 210 days since discovery.   We used the 104 cm Sampurnanand Telescope (ST) \citep{1999CSci...77..643G} and 201 cm Himalayan Chandra Telescope (HCT) \citep{2010ASInC...1..193P} equipped with the broadband {\it BVRI} filters for the photometric monitoring. The images were pre-processed and cleaned using standard IRAF\footnote{IRAF stands for Image Reduction and Analysis Facility distributed by the National Optical Astronomy Observatories which is operated by the Association of Universities for research in Astronomy, Inc., under cooperative agreement with the National Science Foundation.} packages.  In all the images psf (point spread function) photometry was done to extract the instrumental magnitudes following the steps described in detail in \cite{2018MNRAS.474.2551S}.   Given the location of the SN in the host,  the SN flux was contaminated by the host galaxy flux.  To estimate the true SN flux free of any host galaxy contamination we performed template subtraction.   The templates were observed on February 26, 2017 with 201 cm HCT under good photometric conditions.  One such template subtracted image is shown in figure \ref{fig:template}. The instrumental SN magnitude is then measured in the subtracted images. 

To calibrate the SN instrumental magnitude,  we observed three Landolt equatorial standards \citep{2009AJ....137.4186L} PG 0918+029, PG 0942-029 and PG 1525-071 and the SN field on February 26, 2017 under good photometric conditions with seeing $\sim$ 2 arcsec in the {\it V} band with the 201cm HCT.  The brightness of the standard stars was in the range 16.4~\textless~$V$ \textless~12.27 and the colour range was -0.271~\textless~$B-V$ \textless~1.109.  The standard fields were observed at different airmasses ranging between 1.6 and 1.2.  The standard magnitudes and the instrumental magnitudes of the Landolt stars were used to simultaneously fit for the zero points and the colour coefficients following the least square regression technique \citep{1992JRASC..86...71S}.    Since the Landolt fields were observed only for estimating the zero points and the colour coefficients,  we used the site extinction values \citep{2008BASI...36..111S} to correct for the atmospheric extinction.  The coefficients thus obtained are used to transform the instrumental magnitudes of the Landolt stars to standard magnitudes.  Our calibration shows that the rms scatter between the transformed and standard magnitudes is $\sim$0.06 in {\it B} and {\it V} bands and  $\sim$0.03 in {\it R} and {\it I} bands.   Using these transformation equations we generated seven non-variable local standard stars in the SN field (shown in figure \ref{fig:id_figure} and tabulated in Table \ref{tab:standard_star_table}). Using the secondary standards night to night zero-points were determined and SN magnitudes were calibrated differentially.  The resultant error in the SN magnitude is obtained by adding in quadrature the photometric and the calibration errors.  The calibrated SN magnitudes along with the associated errors are listed in Table \ref{tab:photometric_observational_log}.   

Long slit low resolution spectroscopic data were obtained at 27 epochs from 201 cm HCT. We used two grisms Gr7 (3800 - 7800 \AA, resolution 1330) and Gr8 (5800 - 9200 \AA, resolution 2190) in order to cover the visible region. Arc lamps (FeAr and FeNe) and spectrophotometric standard stars (Feige 34, Feige 110 and Hz 44) were observed each night along with SN for wavelength and flux calibration respectively. Necessary preprocessing and extraction of spectra were done under IRAF environment.  The dispersion relations were obtained using the spectra of arc lamp. The dispersion correction was applied to the SN spectra, and OI lines at 5577 \AA, 6300 \AA~and 6364 \AA~were used to cross check the wavelength calibration. In some cases wavelength shift between 0.2 \AA~to 3 \AA~ was found and corrected. The wavelength corrected spectra were then flux corrected using the observed spectrophotometric standard. The flux calibrated  spectra in two grisms Gr7 and Gr8 were combined by applying a scale factor. After wavelength and flux calibration all relative flux spectra were brought to an absolute flux scale using {\it BVRI} photometry.  Log of spectroscopic observation is given in Table \ref{tab:spectroscopic_observations}.    

\section{Characteristics of M12045}\label{sec:Characteristics of M12045} 
\label{sec:characteristics}
M12045 was discovered by MASTER-Tunka auto-detection system \citep{2010AdAst2010E..30L} on 28 October, 2014 22:04 UT  in NGC 4080 at an unfiltered magnitude of 13.9 mag  \citep{2014ATel.6634....1G}.  It was classified as a Type Ib SN, with the presence of well developed \ion{He}{I}, \ion{Fe}{II} and \ion{Ca}{II} features, a few weeks after maximum \citep{{2014ATel.6639....1S}, {2014ATel.6641....1T}}.  Bright radio emission associated with the SN was reported by \cite{2014ATel.6712....1K} and \cite{2014ATel.6755....1C} with the VLA and GMRT, respectively. However no X-ray and UV emission was detected in the {\it Swift} observations \citep{2014ATel.6719....1M}. The basic properties of M12045 and the host galaxy NGC 4080 are listed in Table \ref{tab:NGC4080_detail}.

\begin{table}
\caption{ Details of MASTER OT J120451.50+265946.6 and it's host galaxy NGC 4080.}
\centering
\smallskip
\begin{tabular}{c c }
\hline \hline
Host galaxy & NGC 4080    \\
Galaxy type & SAc C $^\dagger$\\  
Constellation & Virgo  \\ 
Redshift & 0.001891 $^\ddagger$\\ 
Major Diameter & 1.50 arcmin \\
Minor Diameter & 0.70 arcmin \\
Helio. Radial Velocity &  567$\pm$4 km/sec \\
R.A.(J2000.0) & 12$^h$04$^m$51.50$^s$ \\
Dec.(J2000.0) & +29$^d$59$^m$46.6$^s$ \\
Distance modulus  & 31.43 mag \\
Galactic Extincion E(B-V) & 0.02 mag \\
SN type & Ib\\
Offset from nucleus & 4$^{''}$ W,13$^{''}$ N \\
Date of Discovery & 2456959.375 (JD) \\
Estimated date of explosion & 29 September, 2014\\
Estimated date of explosion  (JD) & 2456929.59 \\
Time of maximum in B band & 2456938.49$\pm$2 (JD) \\
Time of maximum in V band & 2456948.91$\pm$3(JD) \\
\hline 
\end{tabular}

$^\dagger$From \cite{2015ApJS..217...27A} 
$^\ddagger$From \cite{1990ApJS...72..245S}                          
\label{tab:NGC4080_detail}      
\end{table}

To constrain the epoch of maximum light/explosion, we used the spectral identification code SNID \citep{2007ApJ...666.1024B} and  GEneric cLAssification TOol (GELATO) \citep{2008A&A...488..383H} on the first spectrum of M12045 obtained on 29 October, 2014.   On the basis of spectral cross correlations with a library of spectral templates, we estimate the explosion epoch to be 29 September, 2014 (JD=2456929.59). The estimated time of maximum in {\it B} and {\it V} bands is found to be between 5--9 October, 2014 (JD = 2456938.49$\pm$2) and 15--21 October, 2014 (JD = 2456948.91$\pm$3) respectively. In Table \ref{tab:snid_gelato_fit} we report the best three matches along with the quality of fit parameter. We have adopted this epoch of $B_{max}$ for further analysis in the paper.

\begin{table}
\caption{The best three matches to the spectrum obtained on 29 October, 2014 which are adopted for age estimation of M12045.}
\centering
\smallskip
\begin{tabular}{c c c c }
\hline \hline
	SNID & SN   & rlap          & Age since V$_{max}$ 		  \\ 
	     &      &               &  (days) 		  \\ 
\hline
	& SN 1998dt       & 12.03   & 7.90  		  \\
	& SN 1990I        & 10.58   & 11.40                 \\
	& SN 1999ex       & 9.17    & 9.30                  \\

 \hline
	GELATO & SN   & Quaity of fit           & Estimated phase \\             
	       &      &  (QoF)                  & (days)\\             
	\hline 
	   & SN 1999dn         & 2.58            & 21.0 (since $B_{max}$) \\
	   & SN 2008ax         & 2.45            & 30.7 (since explosion)   \\
	   & SN 2008ax         & 2.31            & 29.0 (since explosion)   \\
\hline                                   
\end{tabular}
\label{tab:snid_gelato_fit}      
\end{table}

The host galaxy NGC 4080 of M12045 has three consistent distance estimates available using the Tully-Fisher method (15.0 Mpc ($\mu$ = 30.88) \citealt{2013AJ....145..101K}, 15.2 Mpc ($\mu$ = 30.91 $\pm$ 0.54) \citealt{2016AJ....152...50T} and 16.3 Mpc ($\mu$ = 31.06 $\pm$ 0.43) \citealt{2014MNRAS.444..527S}).  We adopt an average distance of 15.5 Mpc ($\mu$ = 30.97 $\pm$ 0.98).  

The estimation of physical parameters from photometric data depends upon an accurate estimation of extinction both Galactic and host along the line of sight of the SN.  The Galactic extinction in the direction of M12045 is found to be $E(B-V)$ = 0.02 mag \citep{2011ApJ...737..103S}.  Due to the location of Type Ib/c SNe in dusty star forming galaxies,  they are expected to suffer from higher host extinction \citep{1996AJ....111.2017V,2008MNRAS.390.1527A,2008ApJ...687.1201K}. From the few initial spectra we find no significant NaID line. However,  high resolution spectra are needed to better determine the host galaxy reddening using the NaID line.   \cite{2011ApJ...741...97D} has given an alternate method to determine the host extinction using the colors.  For a sample of Type Ib/c SNe, \cite{2011ApJ...741...97D} estimate a mean host galaxy reddening of 
${E(B-V)}_{host}$ = 0.36$\pm$0.24 for Type Ib/c SNe.  Based on our low resolution spectroscopic observations we cannot determine the host reddening correctly.  We therefore take the host galaxy reddening value from \cite{2011ApJ...741...97D}.  We adopt a total (Galactic + host) reddening of E(B-V) = 0.38$\pm$0.24 mag and R$_v$ = 3.1 \citep{1989ApJ...345..245C} in this work. 
 
\begin{figure*}
	\begin{center}
		\includegraphics[width=\textwidth]{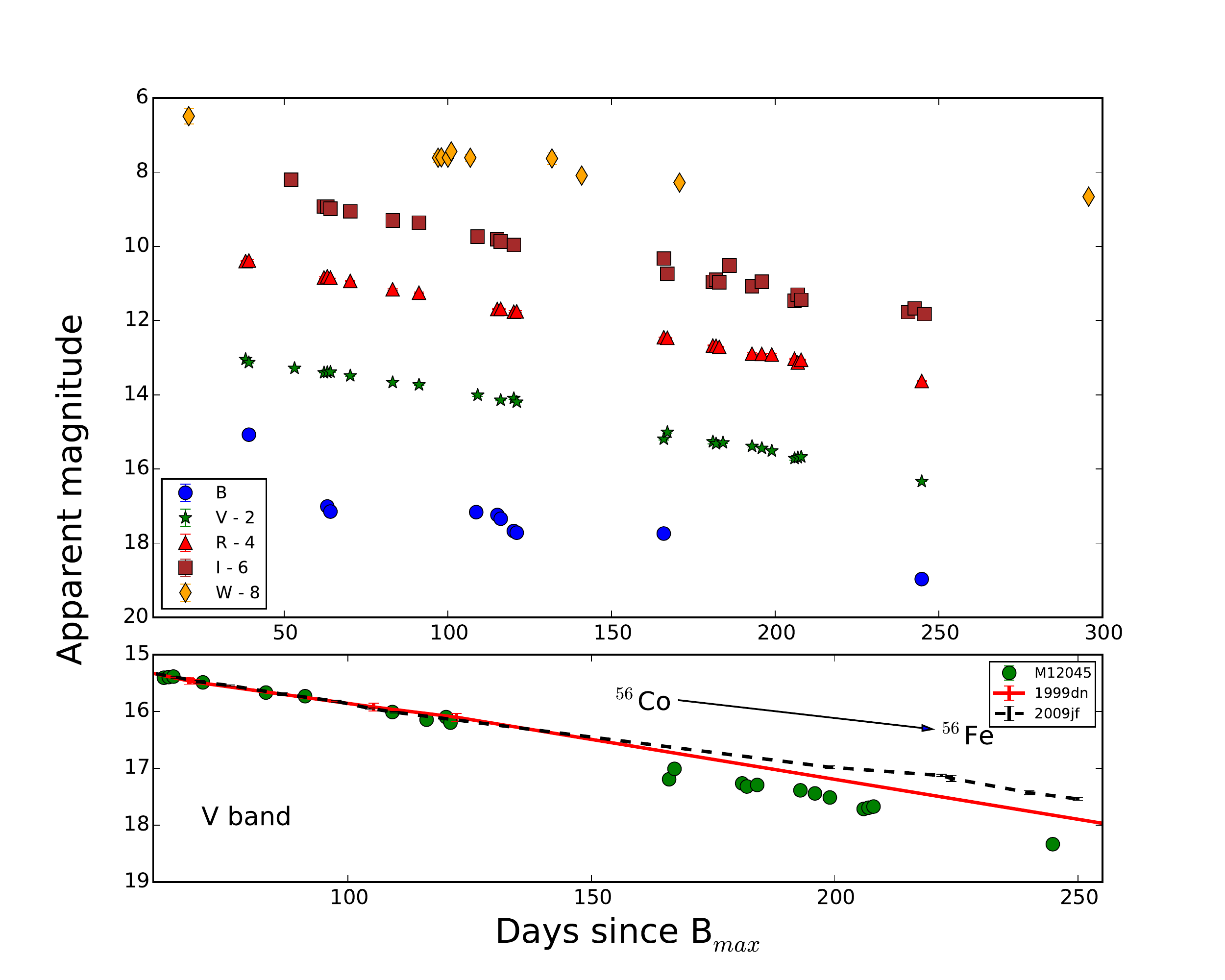}
	\end{center}
	\caption{Top panel:  {\it BVRI} and {\it W} band light curve evolution of M12045. Arbitrary offsets are applied in each band for clarity. Bottom panel:  The $V$ band light curve of M12045 overplotted with those of SNe 1999dn and 2009jf indicating the faster decline of M12045 beyond 150 days since $B_{max}$.}
	\label{fig:light_curve}
\end{figure*}

In Figure \ref{fig:light_curve} we show the {\it BVRI} and {\it W} band light curves of M12045. At discovery the SN was a few weeks past maximum. Our observations started $\sim$17 days after discovery and capture the declining light curves of M12045.  We estimate the decline rates between 63 to 245 days since $B_{max}$ to be 2.29$\pm$0.16, 2.45$\pm$0.22 and 2.50$\pm$0.15 mag 100 d$^{-1}$ in {\it V}, {\it R} and {\it I} bands respectively. The decay rate in {\it B} band is found to be 0.76$\pm$0.31 mag 100 d$^{-1}$ at interval between 63 to 166 days since $B_{max}$. At late phases the light curves of Type Ib SNe are powered by the radioactive decay of $^{56}$Co$\rightarrow$$^{56}$Fe. The decay rates of M12045 light curves are steeper than the standard $^{56}$Co$\rightarrow$$^{56}$Fe decay rates (0.0098 mag day$^{-1}$) implying an optically thin ejecta with inefficient gamma ray trapping. The observed late phase decline rates of M12045 are higher than those found for SNe 1999dn \citep{2011MNRAS.411.2726B} and 2009jf \citep{2011MNRAS.413.2583S} beyond 150 days since $B_{max}$  (Figure \ref{fig:light_curve}; bottom panel).

We construct the quasi bolometric light curve of M12045 by integrating the fluxes between $B$ and $I$ bands.  The integrated fluxes at each epoch are converted to luminosity by adopting a distance of 15.5 Mpc. Since at late phases the contribution from the UV bands is negligible, we only correct for the contribution from the IR bands adopting the numbers given in \cite{2016MNRAS.458.2973P}. \cite{2003ApJ...593..931M}
and \cite{2015ApJ...806..191Y} have given simplified assumptions regarding the fraction of energy deposited due to $\gamma$ rays in the SN ejecta.   Most of the energy in the ejecta is assumed to be generated from $^{56}$Co$\rightarrow$$^{56}$Fe radioactive decay.    We fit the late phase bolometric light curve of M12045 using the expression given by \cite{2015ApJ...806..191Y} described below

\begin{equation}
L(t) = M_{56Ni}\left((S_{56Ni} + S_{56Co})\times(1-e^{-\tau}) + S_{56Co}.f_{p}\right)
\end{equation}

\noindent
where $M_{56Ni}$ is the mass of $^{56}$Ni ejected during the explosion and $f_p$ is the positron fraction.  $S_{56Ni}$ and $S_{56Co}$ are the energy release rates expressed as 

\begin{equation}
S_{56Ni} = \left(3.90\times10^{10}\right)e^{-t/t_{56Ni}}~~ {\rm erg~~ s^{-1} g^{-1}}
\end{equation} 

\noindent
and

\begin{equation}
S_{56Co} = \left(7.10\times10^{9}\right)\times\left(e^{-t/t_{56Co}}-e^{-t/t_{56Ni}}\right)~~ {\rm erg~~ s^{-1} g^{-1}}
\end{equation} 

\noindent
where $t_{56Ni}$ = 8.8 and $t_{56Co}$ = 113.5 days are the decay time scales.  The parameter $\tau$ is defined by 

\begin{equation}
\tau = 1000\left(M_{ej}^{2}/M_{\odot}\right) \times \left(E_{k}^{-1}/10^{51}~ {\rm erg ~~s^{-1}} \right) {(t/day)}^{-2}
\end{equation}

\noindent
where $M_{ej}$ is mass and $E_{k}$ is the kinetic energy of the ejecta. 

We adopt $\tau$ = 5.4 to fit the late phase bolometric light curve of M12045 and estimate $^{56}$Ni to be 0.13 M$_\odot$\footnote {If we consider only Galactic reddening the estimated $^{56}$Ni is 0.09 M$_\odot$.}.  A fast declining light curve indicates towards a fainter bolometric light curve, early commencement of $\gamma$ ray escape and highly mixed $^{56}$Ni \citep{1994ApJ...429..300W}. 

\section{Key Spectroscopic Features}\label{sec:prime spectral features}
\begin{figure}
	\begin{center}
		\includegraphics[scale=0.48]{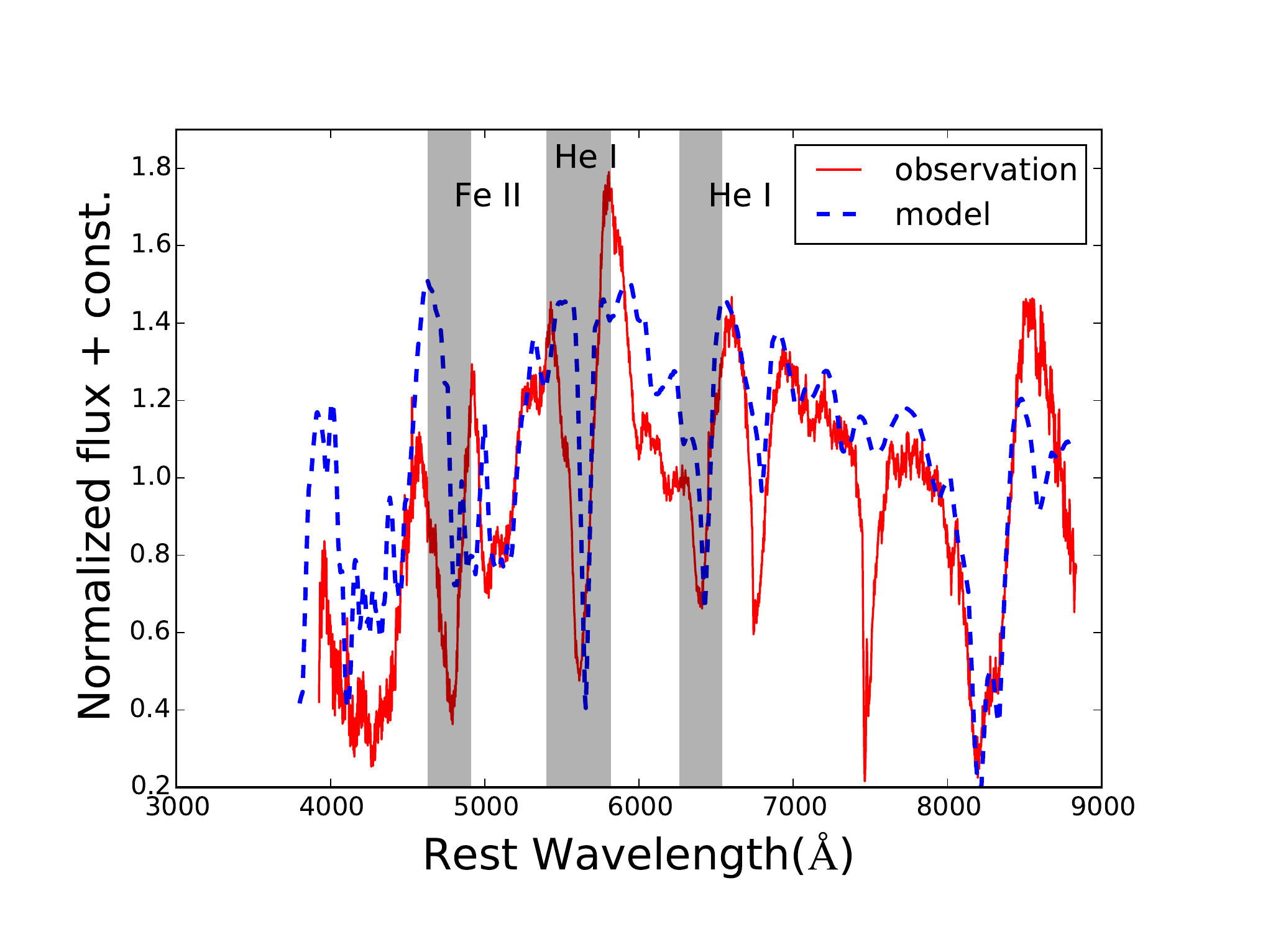}
	\end{center}
	\caption{Early spectrum of M12045 along with the SYN++ spectral fitting.}
	\label{fig:classification_spectrum}
\end{figure}

\begin{figure}
	\begin{center}
		\includegraphics[scale=0.42]{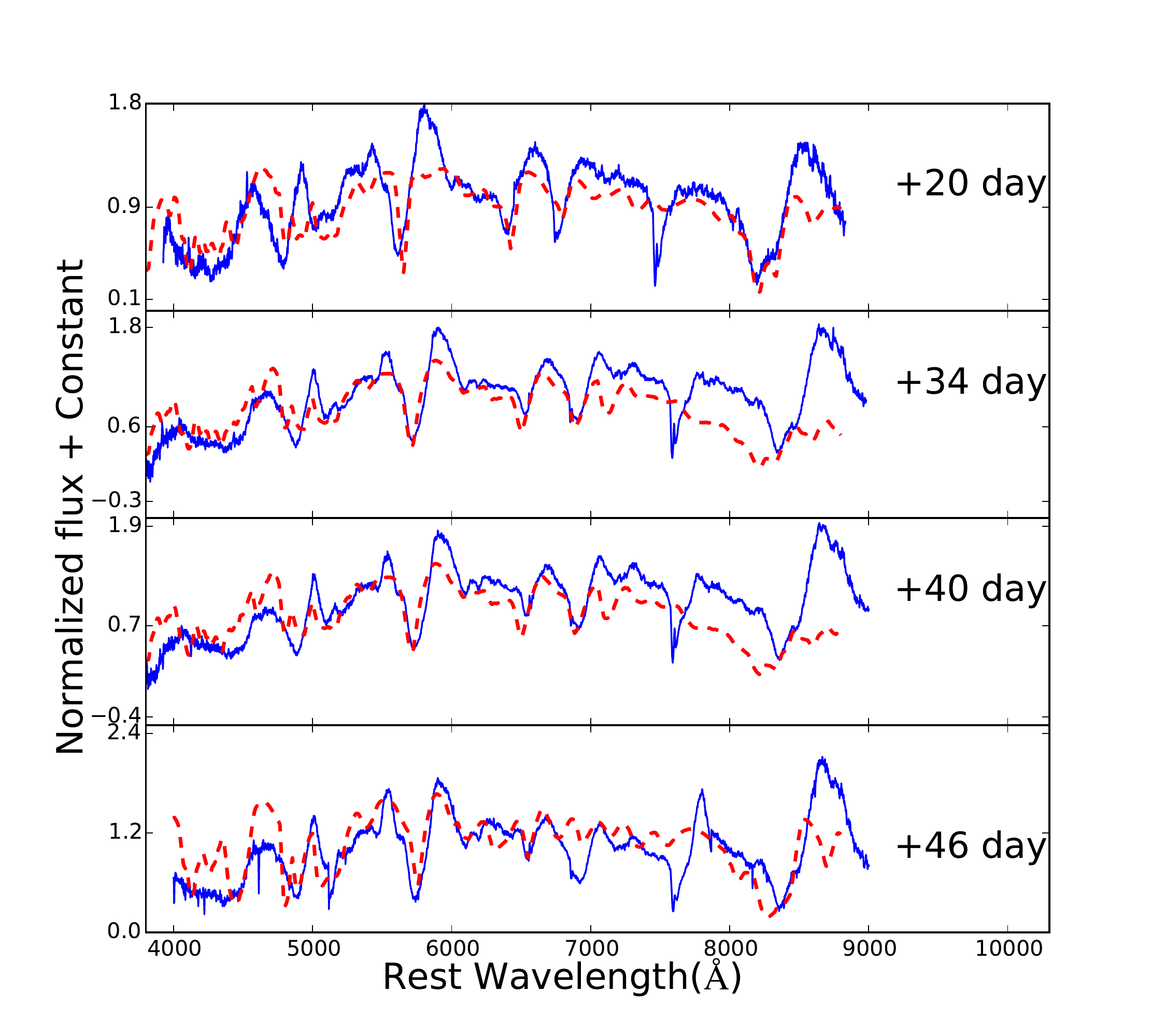}
	\end{center}
	\caption{Same as figure \ref{fig:classification_spectrum} and at later epochs.}
	\label{fig:syn++_fitting}
\end{figure}

In Figure \ref{fig:classification_spectrum} we present the first spectrum of M12045 taken on 29 October, 2014 ($\sim$20 days since $B_{max}$, \citep{2014ATel.6639....1S}). The spectrum is characterized by broad P Cygni profile of \ion{He}{I}, \ion{Fe}{II} etc.  In Figure \ref{fig:classification_spectrum} we also show the best matching synthetic spectrum obtained by SYN++ \citep{2007AIPC..924..342B,2011PASP..123..237T}.  The absorption features due to \ion{He} {I},  \ion{Fe}{II} multiplet and \ion{Ca}{II} NIR are well reproduced in the synthetic spectrum. The photospheric velocity associated with the best fit spectrum is 7500 km s$^{-1}$ with a blackbody temperature of 5000 K.  The Boltzmann excitation temperature of various lines varies between 10000K to 20000K.
The spectra fitting is done at four other epochs (20, 34, 40 and 46 days since $B_{max}$) and shown in Figure \ref{fig:syn++_fitting}. 
A gradual decrease in the fit parameters such as velocity and temperature is expected at later phases.  The photospheric velocity during this phase varies between 7500 km s$^{-1}$ to 6000 km s$^{-1}$ whereas the blackbody temperature varies between 5000 K to 4000 K.   Due to the blackbody approximation used in SYN++,  it is not possible to fit the spectra at very late phases.      

\subsection {Spectral Evolution}\label{sec:evolution}
\begin{figure*}
	\begin{center}
		\includegraphics[width=\textwidth]{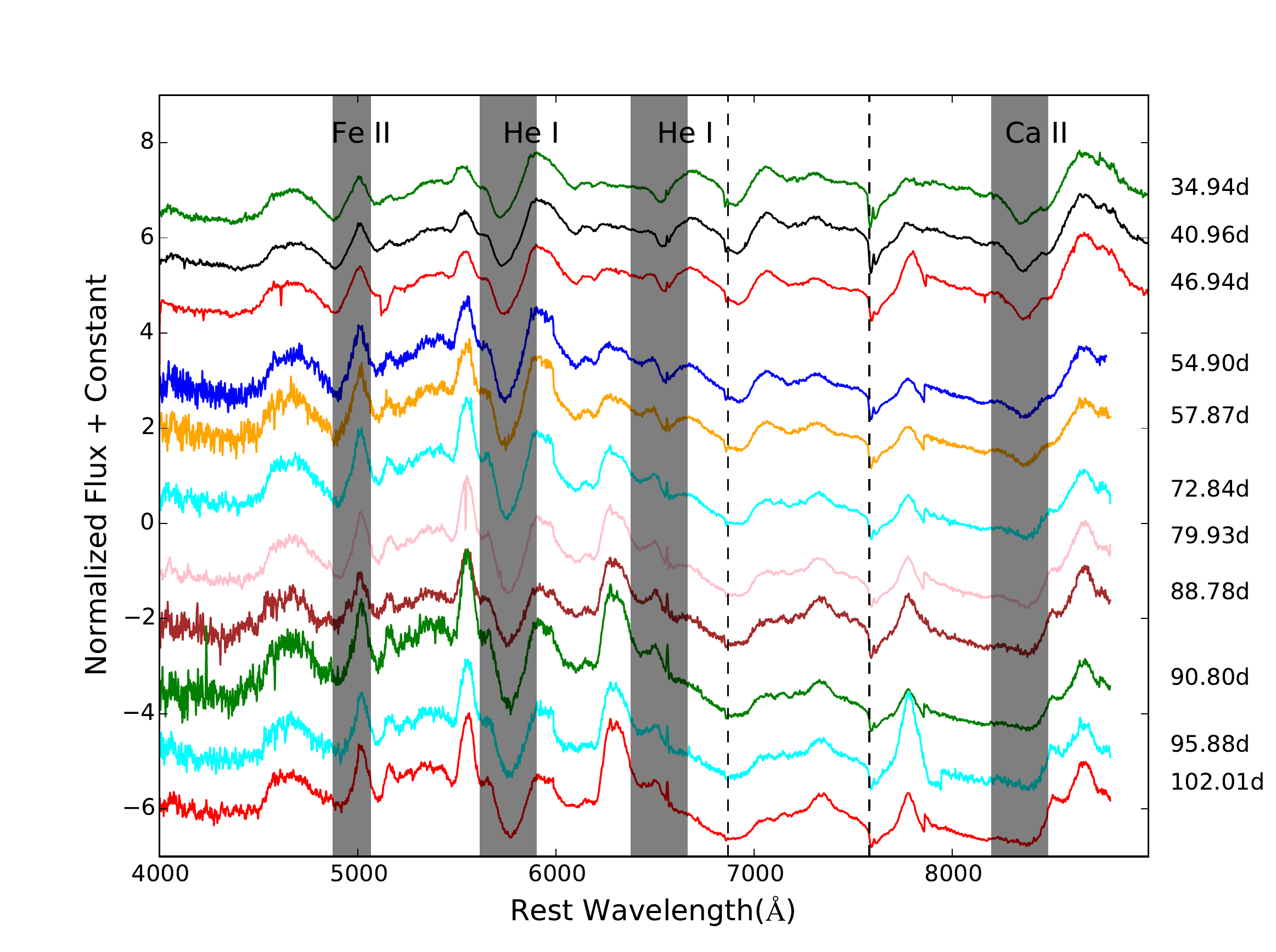}
	\end{center}
	\caption{Spectral sequence of M12045 during 34 to 102 days since $B_{max}$. Prime spectral features are shown with shaded regions. Telluric features are also marked in this figure by dashed line.}
	\label{fig:spectra1}
\end{figure*}

\begin{figure*}
	\begin{center}
		\includegraphics[width=\textwidth]{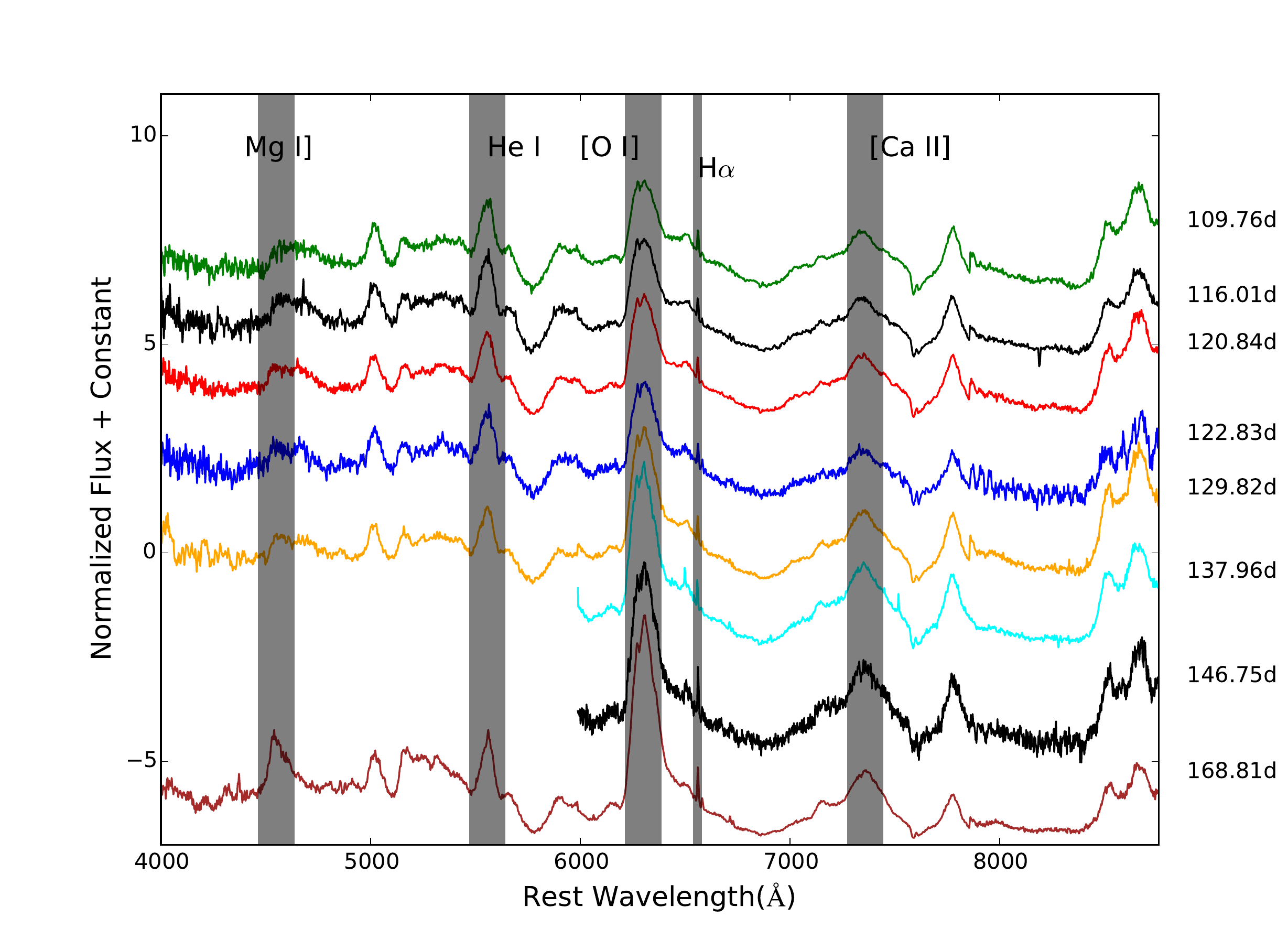}
	\end{center}
	\caption{Spectral sequence of M12045 during 109 to 168 days since $B_{max}$.}
	\label{fig:spectra2}
\end{figure*}

\begin{figure*}
	\begin{center}
		\includegraphics[width=\textwidth]{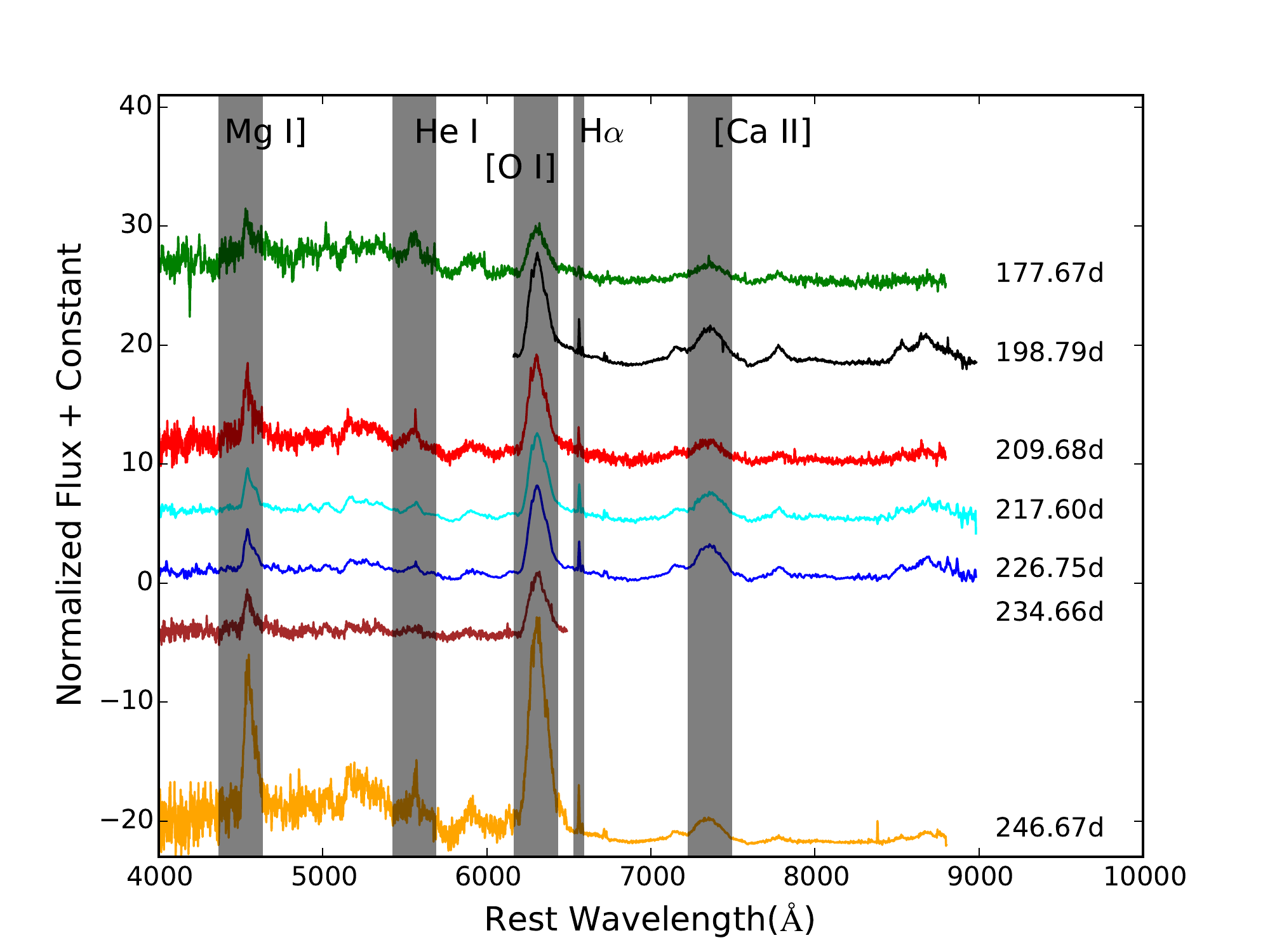}
	\end{center}
	\caption{Nebular phase spectra of M12045 during 177 to 246 days since $B_{max}$. This spectral sequence is dominated by emission features of various lines indicated by shaded regions.}
	\label{fig:spectra3}
\end{figure*}

The spectral evolution of M12045 during early to late nebular phase ($\sim$ 246 days since discovery) is shown in Figures \ref{fig:spectra1}, \ref{fig:spectra2}, \ref{fig:spectra3}. The spectral evolution of M12045 is similar to normal Type Ib SNe.  The prominent and noticeable identifying feature of Type Ib SNe is the \ion{He}{I} line at 5876 \AA~which is clearly seen in the spectra of M12045. The other two He features at 6678 \AA~and 7065 \AA~ are not as prominent as He 5876 \AA~ feature. \ion{Fe}{II} feature near 5000 \AA~ is also prominent in the spectral sequence upto $\sim$ 168 days since $B_{max}$. The absorption signature of \ion{He}{I} line is stronger than any other absorption feature in the spectral evolution. The absorption trough of \ion{Ca}{II} NIR triplet is clearly present in the spectral sequence upto $\sim$ 95 days since $B_{max}$ and is very weak thereafter.  

\begin{figure}
	\begin{center}
		\includegraphics[scale=0.44]{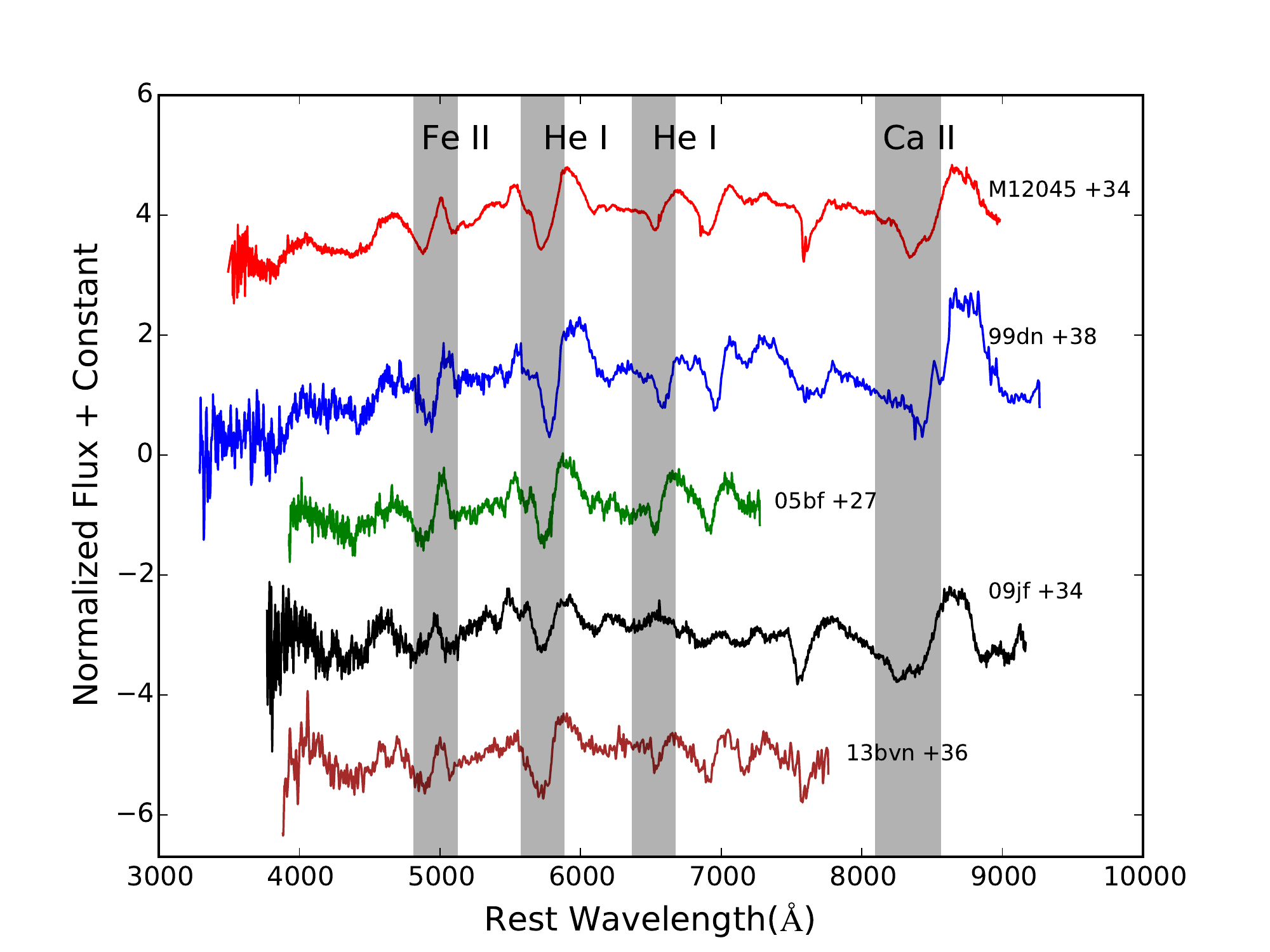}
	\end{center}
	\caption{Comparison of spectral features of M12045 at 34 days since B$_{max}$ with other well studied Type Ib SNe. Prominent He feature is present in all the SNe shown here.}
	\label{fig:34days_spectra_comparison}
\end{figure}

To investigate the spectroscopic behaviour of M12045, we have compared the spectral features with other well studied Type Ib SNe. Figure \ref{fig:34days_spectra_comparison} presents early phase (34 days since B$_{max}$) spectral comparison of M12045 with other Type Ib events such as SNe 1999dn \citep{2011MNRAS.411.2726B}, 2005bf \citep{2005ApJ...633L..97T,2008ApJ...687L...9M,2014AJ....147...99M}, 2007Y \citep{2009ApJ...696..713S}, 2009jf \citep{2011MNRAS.413.2583S} and iPTF13bvn \citep{2014MNRAS.445.1932S}.
 
The spectrum of M12045 is very similar to those of other Type Ib SNe. The  P-Cygni profiles of \ion{Fe}{II} and \ion{He} {I} lines match very well in all the SNe presented in the Figure \ref{fig:34days_spectra_comparison}. The velocity estimated using \ion{Fe}{II} and \ion{He}{I} absorption  are 9000 km s$^{-1}$ and 13000 km s$^{-1}$, respectively at this epoch. \ion{Fe}{II} profile of M12045 matches well with SN 2009jf whereas in the case of SNe 1999dn and 2005bf it is a double peaked structure and in the case of iPTF 13bvn it is a multipeaked profile. He I P-Cygni profile also follows same trend. \ion{He}{I} feature near 6678 \AA~ is present in all the SNe in Figure \ref{fig:34days_spectra_comparison} except SN 2009jf. 

Figure \ref{fig:spectra2} presents the spectral evolution from 109 to 168 days since $B_{max}$. As the spectra evolves absorption features of various lines gradually disappear and the emission features become prominent. In the spectral evolution of M12045, [\ion{O}{I}] doublet at 6300 and 6363 \AA~ starts appearing after $\sim$ 79 days since $B_{max}$ , and is well developed  after $\sim$ 88 days since $B_{max}$. The profile of [\ion{O}{I}] line is initially flat topped which could be because of the blending of the 6300 and 6363 \AA~ lines. In the later epochs the profile of [\ion{O}{I}] lines appears to be asymmetric with bluer component suppressed.  Other dominating emission features at nebular phase are \ion{Mg}{I}] at 4571 \AA~ and [\ion{Ca}{II}] 7291, 7324 \AA. There is possibility of blending of this emission feature with [\ion{Fe}{II}] lines at 7155 \AA, 7172 \AA, 7388 \AA~ and 7452 \AA. The [\ion{Ca}{II}] can also get blended with [\ion{O}{II}] emission line at 7320 \AA~ and 7330 \AA.

\begin{figure}
	\begin{center}
		\includegraphics[scale=0.44]{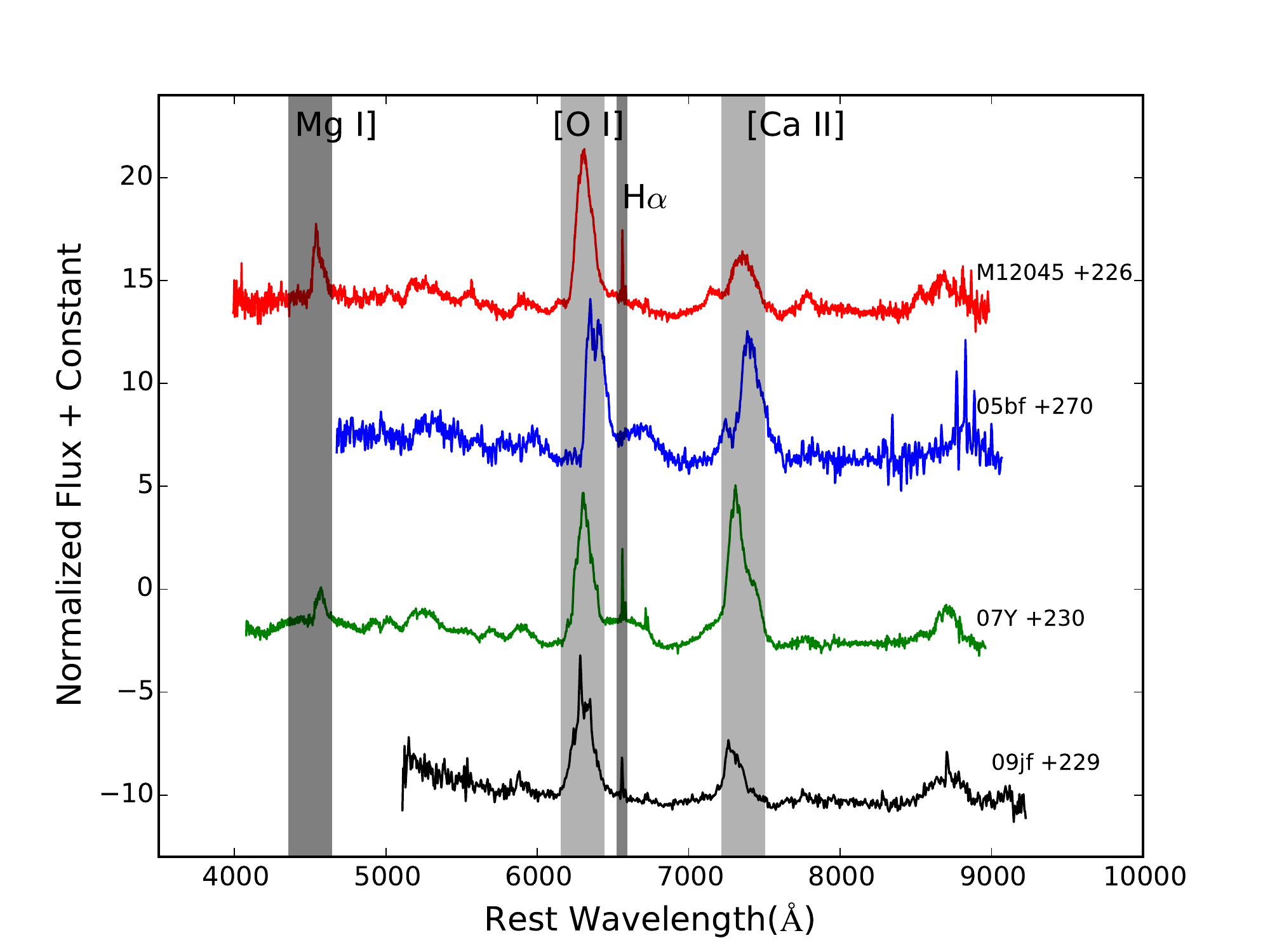}
	\end{center}
	\caption{Comparison of spectral features of M12045 at 226 days since B$_{max}$ with other well studied Type Ib SNe. Due to limited wavelength coverage blue spectral regime is not available for SNe 2005bf and 2009jf. The [\ion{O}{I}] line profile in M12045 shares a similarity with SN 2007Y. [\ion{Ca}{II}] doublet is clearly present in M12045 as well as in the comparison SNe.}
	\label{fig:226days_spectra_comparison}
\end{figure}

Figure \ref{fig:spectra3} presents nebular phase spectral sequence during $\sim$ 177 to $\sim$ 246 days since $B_{max}$. At these very late epochs ejecta becomes optically thin and the deeper layers of the ejecta can be probed. The spectral sequence during late nebular phase is mostly dominated by forbidden emission features of \ion{Mg}{I}], [\ion{O}{I}], [\ion{Ca}{II}].   Nebular lines are basically representative of one dimensional line of sight projection for three dimensional distribution of elements, and so contains vital information about the  geometry of the ejecta \citep{2008Sci...319.1220M, 2008ApJ...687L...9M,2009MNRAS.397..677T}. There are some narrow lines also seen in these spectra, which could be  identified as H$\alpha$, [\ion{N}{II}] 6548, 6538 \AA~and [\ion{S}{II}] 6717, 6731 \AA~features originated from the underlying \ion{H}{II} region \citep{2011MNRAS.413.2583S}.  Figure \ref{fig:226days_spectra_comparison} presents the spectral comparison of M12045 at 226 days since $B_{max}$. The profile of [\ion{O}{I}] doublet is found to be different in the spectra of objects used for comparison. The profile of [\ion{O}{I}] doublet of M12045 shares similarity with SN 2007Y. Whereas SNe 2005bf and 2009jf have clear horned shape/double peaked structure of [\ion{O}{I}] doublet. The double-peaked profile could be due to the explosion geometry. This can also be interpreted as superposition of oxygen lines with high velocity H$\alpha$ feature \citep{2010MNRAS.409.1441M}. The relative strength of [\ion{O}{I}] to [\ion{Ca}{II}] line is found to be different in these spectra. The ratio of  [\ion{O}{I}] to [\ion{Ca}{II}] line flux is relatively high in case of M12045 and SN 2009jf and low in SN 2005bf and SN 2007Y. 

\subsection { Mg I] line, [O I] doublet}\label{sec:emission lines}
\begin{figure*}
	\begin{center}
			\includegraphics[width=\textwidth]{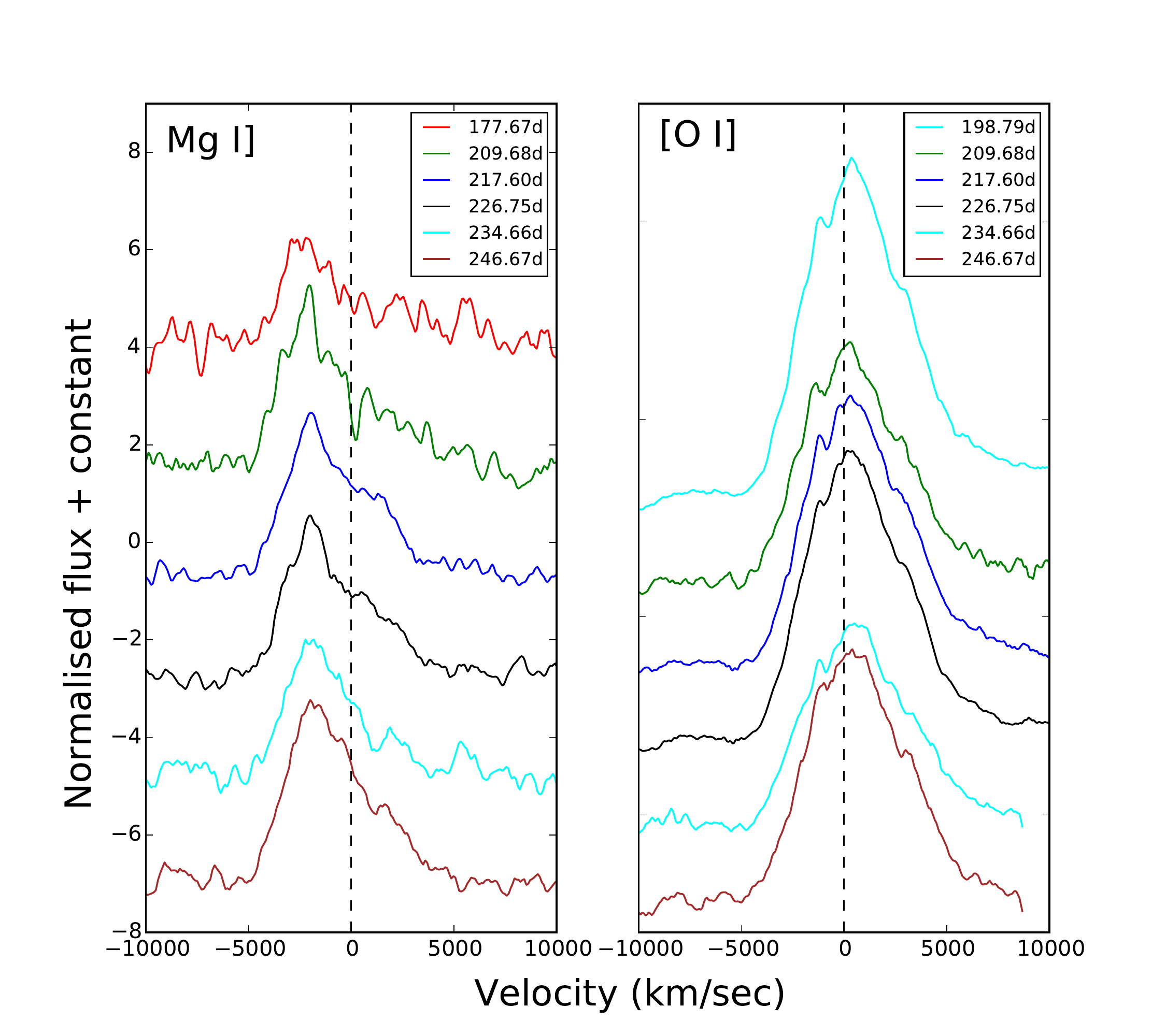}
	\end{center}
	\caption{\ion{Mg}{I}] and [\ion{O}{I}] line evolution in nebular phase of M12045. A prominent difference can be seen in both the profiles in terms of the shift from zero velocity.}
	\label{fig:zoomed_line_velocity_plot}
\end{figure*}

The late nebular phase spectra of stripped envelope SNe are dominated by the forbidden emission lines for eg. [\ion{O}{I}] and  [\ion{Ca}{II}]. The semi-forbidden \ion{Mg}{I}] line (4571 \AA),  which is due to 3s$^{2}$ $^{1}$S$_{0}$ -- 3s3p$^{3}$P$^{0}_{1}$
transition, is prominently seen in the spectra of M12045 obtained after $\sim$ 168 days since B$_{max}$ (Figures \ref{fig:spectra2} and \ref{fig:spectra3}).   
\cite{2006ApJ...640..854M} suggest that Mg and O have similar spatial distribution in the SN ejecta. However, line 
distributions of Mg and O usually deviate from other heavier elements such as Ca or Fe \citep{2005Sci...308.1284M}.  Figure \ref{fig:zoomed_line_velocity_plot} provides a good opportunity to make a comparison between \ion{Mg}{I}] and [\ion{O}{I}] evolution features. In the case of M12045, [\ion{O}{I}] profiles are broader than \ion{Mg}{I}] line profiles which indicates blending with nebular emission lines of other elements. In the present case line profiles are different in terms of the shift from zero velocity. We can see a significant blueshift in the case of \ion{Mg}{I}] line profile whereas [\ion{O}{I}] doublet profile is slightly redshifted. Blueshift of \ion{Mg}{I}] line is because of residual opacity generated from multiple Fe transitions \citep{2009MNRAS.397..677T}. While the slight redshift in the [\ion{O}{I}] profile arises mostly due to the asymmetry in the ejecta.

\subsection {Mass of neutral oxygen and [O I]/[Ca II] ratio}
\label {sec:O I line emission}
Late nebular phase spectra of Type Ib SNe are enriched with [\ion{O}{I}] emission lines. Explosion geometry can be inferred from the [\ion{O}{I}] doublet profile. This feature is fairly isolated and unblended. Other prominent late nebular features as [\ion{Ca}{II}] and \ion{Mg}{I}] are contaminated by different nearby line blending.  In Figure \ref{fig:zoomed_line_velocity_plot} we present evolution of oxygen profile during late phases. As we discussed earlier in section \ref{sec:evolution}, [\ion{O}{I}] doublet profile is asymmetric with the suppression of blue component. In Figure \ref{fig:zoomed_line_velocity_plot} we see a shifted redward component from the zero velocity. These signatures suggest that the explosion of M12045 is axisymmetric. This geometry is linked with oxygen distributed as a torus viewed from the equatorial direction or a blob of oxygen moving perpendicular to the line of sight \citep{2001ApJ...559.1047M,2002ApJ...565..405M,2006ApJ...640..854M}.     

By measuring the flux of the [\ion{O}{I}] line,  we can estimate the mass of neutral oxygen following the expression given by \cite{1986ApJ...310L..35U} as described below. 
 
\begin{equation}
M_{O} = 10^{8} \times D^{2} \times F([\ion{O}{I}]) \times e^{(2.28/T_{4})}
\end{equation}

\noindent
where M$_{O}$ is mass of neutral oxygen in term of M$_{\odot}$, D is the distance to the SN in Mpc, F([\ion{O}{I}]) is the observed absolute flux of [\ion{O}{I}]  line  in erg s$^{-1}$ cm$^{-2}$ and T$_{4}$ is the temperature associated with oxygen emitting region in units of 10$^{4}$ K. The above equation holds good in high density regime (N$_{e}$ $\geq$ 10$^{6}$ cm$^{-3}$), associated with the ejecta of SNe Type Ib \citep{1989AJ.....98..577S,1994AJ....108..195G,2004A&A...426..963E}. Temperature of the line emitting region can be estimated using flux ratio of [\ion{O}{I}] 5577/6300$-$6344 lines. In the observed  spectra of M12045 during the late nebular phase, [\ion{O}{I}] 5577 \AA~line is not clearly detected and hence an upper limit of $\sim$ 0.1 can be set  on the ratio of [\ion{O}{I}] 5577/6300$-$6364 line flux. For this limit, the line emitting region can be either  a low temperature (T$_4\leq$ 0.4), high density  region  or a high temperature region (T$_4$ = 1) with low electron density (n$_e\leq$ 5$\times$10$^6$ cm$^{-3}$  \citealt{2007ApJ...666.1069M}). In the $\sim$ 226 d spectrum,  the observed flux of [\ion{O}{I}] line,   F([\ion{O}{I}]) =  1.25 $\times$ 10$^{-13}$ erg s$^{-1}$ cm$^{-2}$,  distance of 15.5 Mpc, and considering the case of low temperature (T$_4$ = 0.4), the estimated mass of neutral oxygen is 0.90 M$_\odot$. 

The prime cause of [\ion{O}{I}] emission is the formation of a layer of oxygen during the hydrostatic burning phase. The main sequence mass of the progenitor is directly related with the ejected mass of oxygen.  \cite{1996ApJ...460..408T} have predicted nucleosynthesis yields for progenitor masses between 13--25 M$_{\odot}$ based on nucleosynthesis calculations.    Based on the calculations, \cite{1996ApJ...460..408T} showed that for oxygen masses 0.22, 0.43, 1.48, and 3.00 M$_{\odot}$,  the associated progenitor mass should be 13, 15, 20 and 25 M$_{\odot}$ respectively.   For 13, 15 and 25 M$_{\odot}$ progenitor masses,  the corresponding He core mass estimates given by \cite{1996ApJ...460..408T} were 3.3, 4.0 and 8.0 M$_{\odot}$ respectively.   The estimated oxygen mass of 0.90 M$_\odot$ in M12045 indicates towards a main sequence progenitor mass of 15--20 M$_{\odot}$ with a He core mass between 4--8 M$_{\odot}$.

Using the ratio of [\ion{O}{I}] and [\ion{Ca}{II}] line fluxes we can estimate the main sequence mass of the progenitor.   The ratio of the fluxes of [\ion{O}{I}] and [\ion{Ca}{II}] lines is found to be insensitive to density, temperature whereas it increases with an increasing progenitor mass \citep{1989ApJ...343..323F,2004A&A...426..963E}.  In the case of M12045 this line ratio is between 2.1 to 2.5 for spectra between $\sim$168 to 226 days since $B_{max}$.   This line ratio is nearly constant for SNe post $\sim$280 days since explosion \citep{2004A&A...426..963E}.  \cite{2015A&A...579A..95K} have compared this ratio for a number of stripped envelope SNe and found that the line ratio for Type II SNe does not exceed 0.7 while there is a considerable spread seen for Type Ib/c SNe ($\sim$ 0.9 to 2.5).   As stated in \cite{2015A&A...579A..95K},  this spread in line ratio indicates towards two different progenitor populations for Type Ib/c SNe - those coming from massive WR stars and those from lower mass progenitors in binary systems.  The flux ratio of M12045 indicates that it is associated with a massive star progenitor. The line ratios in SNe 1999dn and 2009jf were found to be $\sim$2 and $\sim$1.82, respectively with an associated progenitor mass of $\sim$25 M$_{\odot}$ \citep{2011MNRAS.411.2726B, 2011MNRAS.413.2583S}. The estimated mass of neutral oxygen in M12045 also indicates the progenitor mass between to be 15--20 M$_{\odot}$. Based on the above two estimates we can say that a massive star of main sequence mass $\sim$ 20 M$_{\odot}$ is likely the progenitor of M12045.  

\subsection {Velocity evolution}
\label {sec:Velocity evolution}
\begin{figure*}
	\begin{center}
		\includegraphics[scale=0.8]{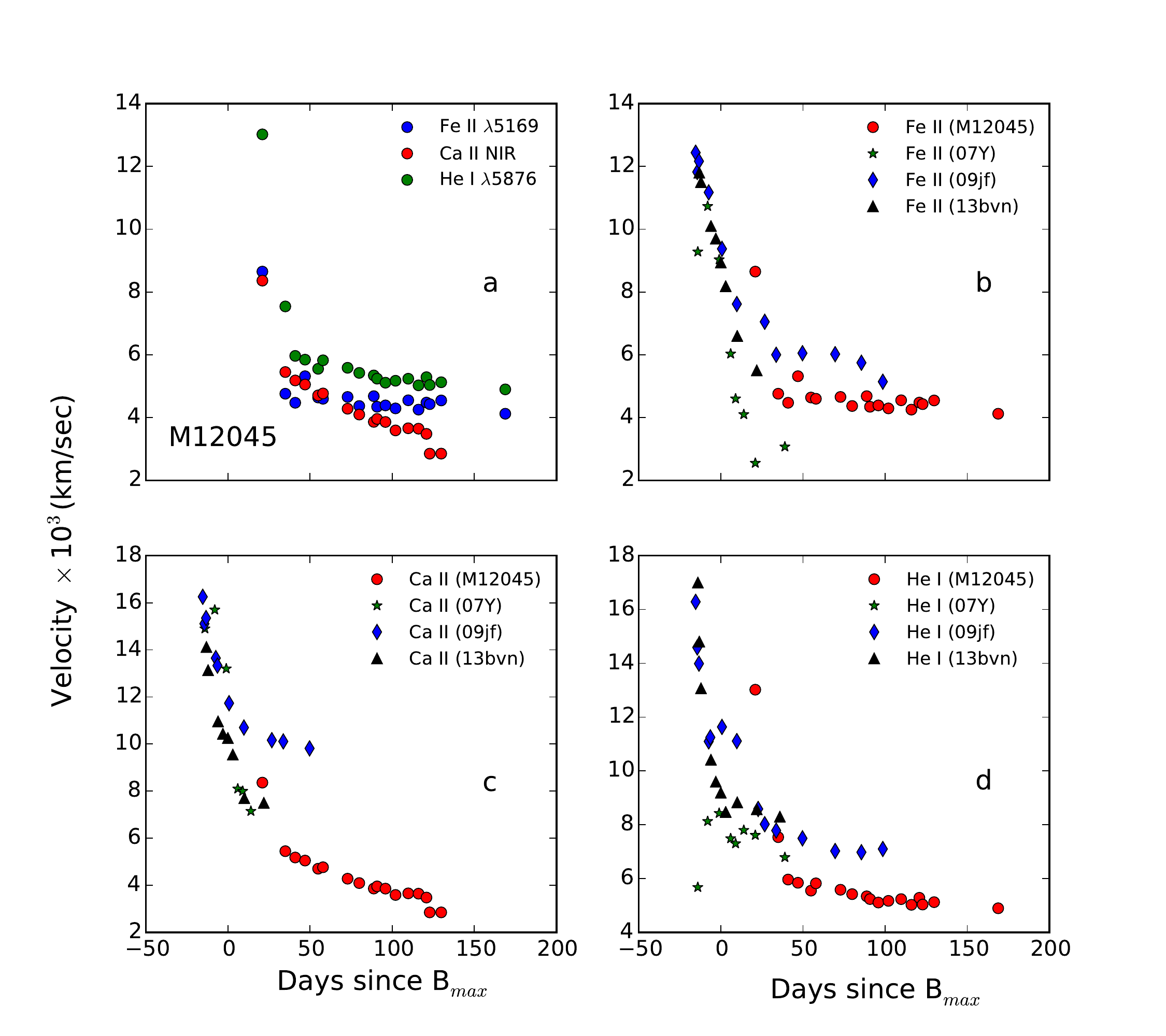}
	\end{center}
	\caption{The temporal evolution of the velocities of \ion{Fe}{II} 5169\AA~, \ion{He}{I} 5876 \AA~ and \ion{Ca}{II} NIR for M12045 are shown in the upper left subplot. For the same lines we show a comparison of velocities of M12045 with SNe 2007Y \citep{2009ApJ...696..713S}, 2009jf \citep{2011MNRAS.413.2583S} and iPTF 13bvn \citep{2014MNRAS.445.1932S} in the remaining three panels.}
		\label{fig:velocity_plot}
\end{figure*}

We estimated velocities of a few lines from the absorption minima of  P-cygni features. Figure \ref{fig:velocity_plot} presents the velocity evolution of \ion{Fe}{II} $\lambda$ 5169 \AA, \ion{He}{I} $\lambda$5876 \AA~and \ion{Ca}{II} NIR triplet in M12045.   A comparison with a few well studied Type Ib SNe is also shown (SNe 2007Y \citealt{2009ApJ...696..713S}, 2009jf \citealt{2011MNRAS.413.2583S} and iPTF13bvn \citealt{2014MNRAS.445.1932S}) in Figure \ref{fig:velocity_plot}. During the first few days the line velocities for M12045 declines rapidly and then attains a relatively constant value. The velocities estimated using \ion{Fe}{II}  5169\AA, \ion{He}{I} 5876 and \ion{Ca}{II} NIR triplet at $\sim$ 20 days since $B_{max}$ are $\sim$ 8700, 13000 and 8300 km s$^{-1}$, respectively.  The velocity of \ion{Fe}{II} line at 5169 \AA~ at $\sim$ 20 days since $B_{max}$ is lower than what we expect at maximum and is consistent with the average photospheric velocities between 8000$\pm$2000 km s$^{-1}$ for Type Ib/c SNe at maximum \citep{2013MNRAS.434.1098C}. At $\sim$ 35 days since $B_{max}$ these velocities drop to $\sim$ 4700, 7500 and 5500 km s$^{-1}$. Beyond $\sim$ 46 days since $B_{max}$  the line velocities of \ion{Fe}{II} and \ion{He}{I} remain nearly constant at $\sim$ 5000 km s$^{-1}$, whereas \ion{Ca}{II} line velocity further declines linearly to $\sim$ 3000 km s$^{-1}$ at 137 days since $B_{max}$. 

Figure \ref{fig:velocity_plot} (b) shows the comparison of \ion{Fe}{II} line at 5169\AA~ with SNe 2007Y, 2009jf and iPTF 13bvn. At $\sim$20 days since $B_{max}$ the velocity of M12045 is higher than all the comparison SNe. At later phases velocity associated with \ion{Fe}{II} line at 5169\AA~ for M12045 is $\sim$ 1500 km s$^{-1}$ lower than that of SN 2009jf and 1700 km s$^{-1}$ higher than velocities associated with SN 2007Y. On the basis of the available data we can see that the velocity of the \ion{Ca}{II} NIR for M12045 is lower than SN 2009jf and higher than SNe 2007Y and iPTF13bvn (Figure \ref{fig:velocity_plot} (c)). At $\sim$20 days since $B_{max}$ the velocity of the \ion{He}{I} line for M12045 is higher than the rest of the comparison sample (Figure \ref{fig:velocity_plot} (d)).  Whereas at later phases the velocity of the \ion{He}{I} line for M12045 seems to be lower than those of other SNe.

\section{Host galaxy metallicity}
\label {sec:Host galaxy metallicity}
In the nebular spectrum of M12045, narrow emission lines are seen (Figure \ref{fig:spectra3}). The observed narrow lines originate from the \ion{H}{II} region in which the SN is embedded.  Using the flux of the narrow emission lines,  we can estimate the metallicity at the SN location in the host galaxy.    There are several diagnostics for metallicity measurements given by several authors \citep{1991ApJ...380..140M,2002ApJS..142...35K,2004MNRAS.348L..59P,2005ApJ...631..231P} which depend upon the various emission lines ratios and calibrations.  However, due to limitation of the  observed  wavelength range ($\sim$ 4000 -- 9000 \AA) and non-detection/very weak  [OIII] lines, we are left with only N2 index calibration of  \cite{2004MNRAS.348L..59P} for estimating the metallicity of the SN region.   In Table \ref{tab:host_galaxy_nebular_lines} we list the flux measurements of nebular emission lines seen in the $\sim$226 d spectrum of M12045.  Based on the N2 index we estimate the metallicity to be 12+log(O/H) = 8.56$\pm$0.18 dex.

\begin{table}
\caption{ Flux measurements of various emission lines in the nebular spectrum ($\sim$ 226 days since $B_{max}$) of M12045.}
\centering
\smallskip
\begin{tabular}{c c c }
\hline \hline
Species        & Wavelength        & Flux                                    \\
               & (\AA)             & (10$^{-17}$)(erg s$^{-1}$ cm$^{-2}$)          \\
\hline
H$\alpha$      & 6563              & 254       \\
NII            & 6583              & 65        \\
SII            & 6717              & 31         \\
SII            & 6731              & 22         \\
\hline                                   
\end{tabular}
\label{tab:host_galaxy_nebular_lines}      
\end{table}

The value of  oxygen abundance 12+log(O/H) for the Sun  available in the literature are  8.69$\pm$0.05 dex (\citealt{2001ApJ...556L..63A}), 8.69 dex (\citealt{2009ARA&A..47..481A}) and  8.76$\pm$0.07 dex (\citealt{2011SoPh..268..255C}). The  oxygen abundance estimated for the SN region is less than the Solar value, indicating marginally sub-Solar metallicity at the SN region within the host galaxy NGC 4080. The sub-Solar metallicity environment of M12045 is consistent with a massive WR progenitor star scenario. The probability which favours a massive WR star as a progenitor system increases with increasing metallicity because stellar winds are metallicity dependent \citep{2010A&A...516A.104L}.

\section{Summary} 
\label{sec:summary}
In the present work we discuss the results obtained for a Type Ib SN M12045 based on the photometric and spectroscopic observations. M12045 is one of the few Type Ib SNe with a rich spectroscopic data set spanning upto $\sim$ 250 days since $B_{max}$.  Our analysis shows that the SN was discovered a few weeks after maximum and we have been therefore able to capture the linearly declining light curves in $BVRI$ bands. The late phase decline rate in all the bands is considerably steeper than the expected $^{56}$Co to $^{56}$Fe  decay indicating an optically thin ejecta with inefficient gamma ray trapping. Fitting a late phase energy deposition function to the bolometric light curve of M12045 yields a $^{56}$Ni mass of 0.13 M$_{\odot}$. Faster decline in the light curve during radioactive tail indicates towards a higher mixing of $^{56}$Ni and hence a smaller $^{56}$Ni production. The spectroscopic evolution of M12045 is similar to a typical Type Ib SN with enriched features of He, O, Mg and \ion{Ca}{II} NIR. The [\ion{O}{I}] doublet profile in the late nebular phase shows asymmetry in the explosion. The spectral sequence of M12045 also shows narrow H$\alpha$ feature originating from the underlying H II region. Photospheric velocity measured from \ion{Fe}{II} line at 5169 \AA~ is consistent with the average velocity of Type Ib/c SNe. The mass of neutral oxygen, the flux ratio of [\ion{O}{I}] and  [\ion{Ca}{II}] lines and sub-Solar metallicity suggests that the progenitor of M12045 could be a massive WR star with a zero age main sequence mass of $\sim$ 20 M$_{\odot}$.  

\section{Acknowledgments}
We thank the observing staff and observing assistants at 104 cm ST and 201 cm HCT for their support during observations of M12045. We acknowledge Wiezmann Interactive Supernova data REPository http://wiserep.weizmann.ac.il (WISeREP) \citep{2012PASP..124..668Y}. This research has made use of the CfA Supernova Archive, which is funded in part by the National Science Foundation through grant AST 0907903. This research has made use of the NASA/IPAC Extragalactic Database (NED) which is operated by the Jet Propulsion Laboratory, California Institute of Technology, under contract with the National Aeronautics and Space Administration. KM acknowledges the support from IUSSTF WISTEMM fellowship and UC Davis. SBP and KM acknowledges BRICS grant DST/IMRCD/BRICS/Pilotcall/ProFCheap
/2017(G) for the present work. BK acknowledges the Science and Engineering Research Board (SERB) under the Department of Science \& Technology, Govt. of India, for financial assistance in the form of National Post-Doctoral Fellowship (Ref. no. PDF/2016/001563). DKS, GCA and BK acknowledge BRICS grant DST/IMRCD/BRICS/PilotCall1/MuMeSTU/2017(G) for the present work. MASTER equipment acknowledges to Lomonosov Moscow State University Development Program. VL acknowledge BRICS grant 17-52-80133. PB work was supported by RNF16-12-00085. SB acknowledges NSFC Project 11573003 and also is partially supported by China postdoctoral science foundation grant No. 2018T110006.

\bibliographystyle{mnras}
\bibliography{refag}

\appendix
\section{Log of photometric and spectroscopic observations}
\begin{figure*}
	\begin{center}
		\includegraphics[width=\textwidth]{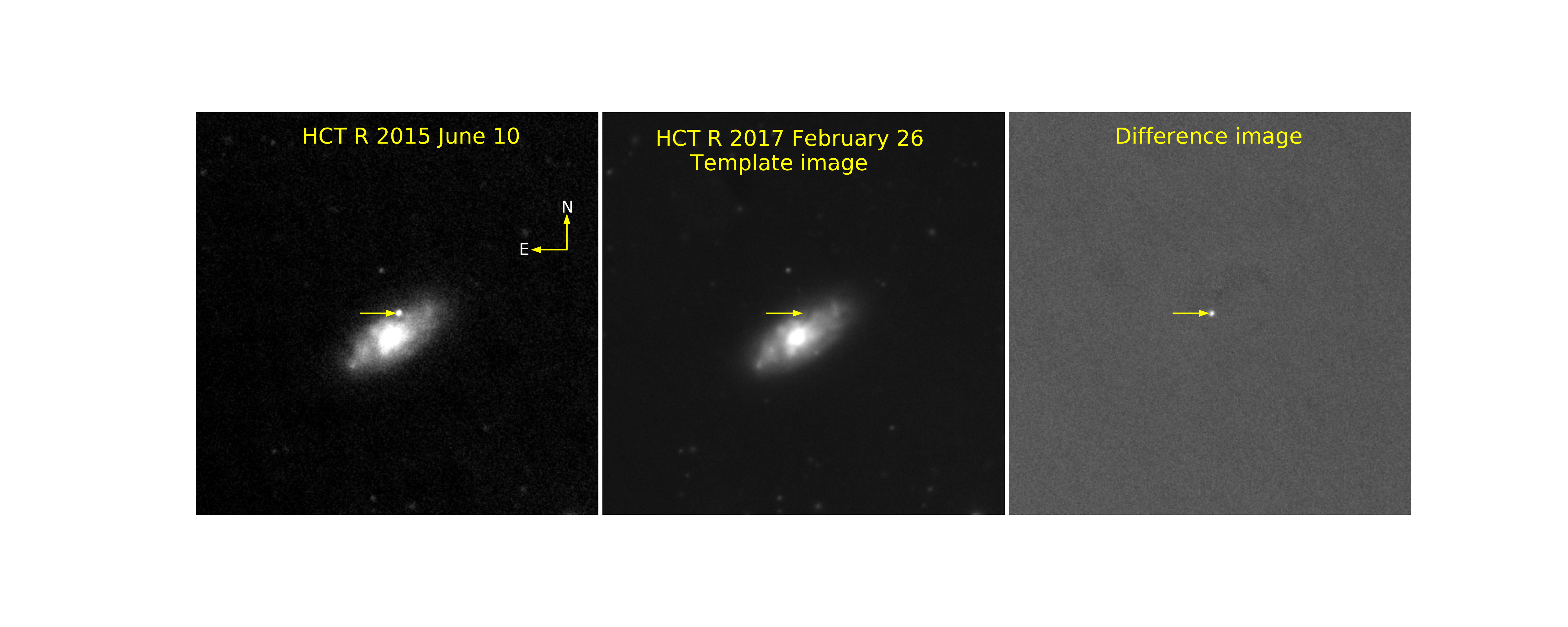}
	\end{center}
\caption{This figure consists of three panels. Left panel is the image acquired with 201cm HCT in R band along with the location of SN marked. Middle panel represents the template image which is obtained on February 26, 2017. Right panel shows the difference image received after template subtraction.}
\label{fig:template}
\end{figure*}

\begin{figure*}
	\begin{center}
		\includegraphics[scale=0.5]{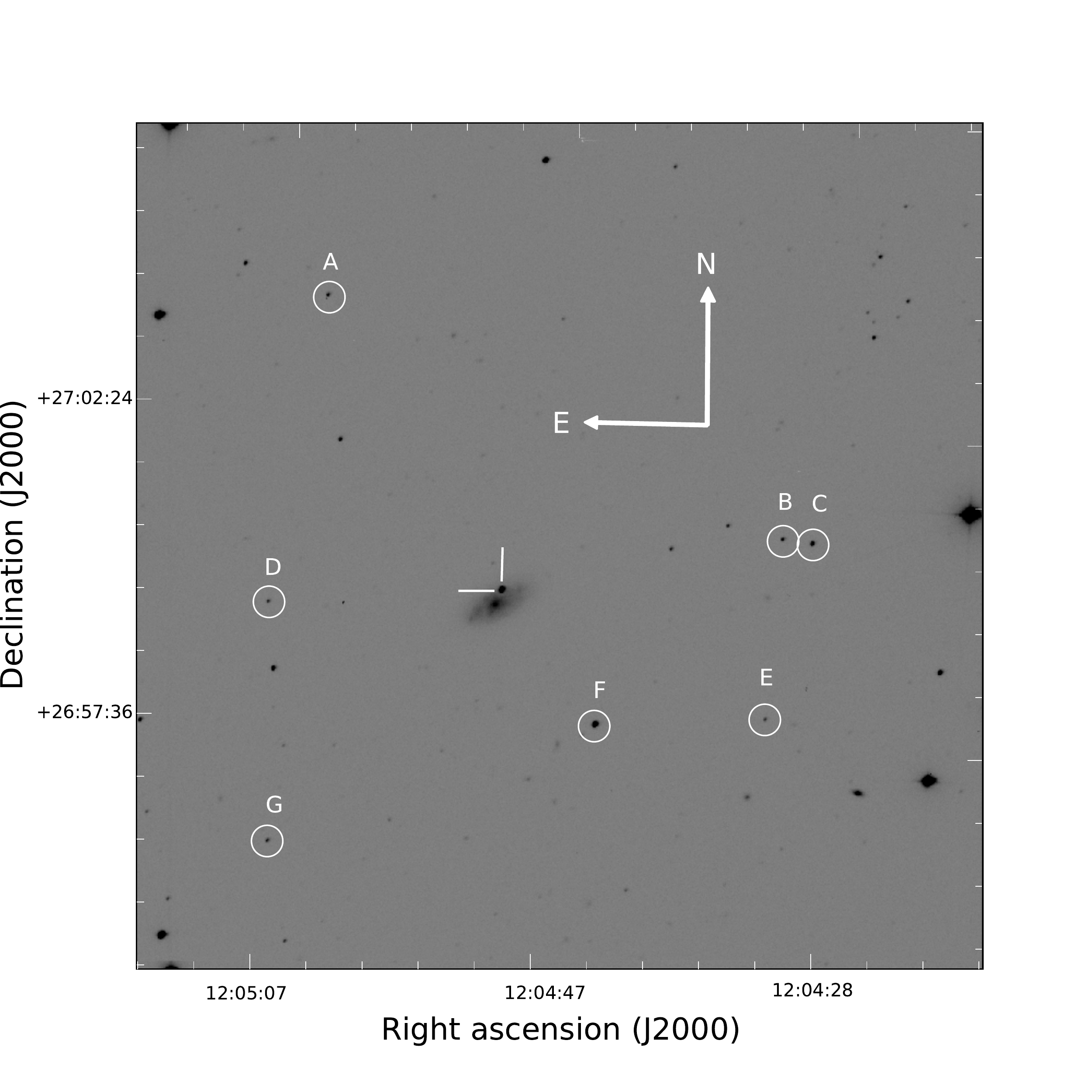}
	\end{center}
\caption{Location of M12045 along with seven local standards used for calibration. The {\it R band} image was taken with 104 cm ST on 15 November, 2014.}
	\label{fig:id_figure}
\end{figure*}

Figure \ref{fig:template} shows one of the template subtracted SN image.   The local standards used for calibration along with the SN location are marked in Figure \ref{fig:id_figure} and listed in Table \ref{tab:standard_star_table}.   The calibrated SN magnitudes and the corresponding errors in {\it BVRI} filters are listed in Table \ref{tab:photometric_observational_log}.    A complete log of spectroscopic observations is given in Table \ref{tab:spectroscopic_observations}. 

\begin{table*}
	\caption{Star ID along with co-ordinates and magnitude of secondary standard stars in { \it BVRI} bands.}
\centering
\smallskip
\begin{tabular}{c c c c c c c }
\hline \hline
Star ID  & R.A.(h:m:s) & Dec.(d:m:s)      &  {\it B}(mag)               &       {\it V}(mag)                  &      {\it R}(mag)                   &      {\it I}(mag)               \\
                                      
\hline
A        & 12:05:04.52  & +27:04:06.8     &  17.882 $\pm$ 0.004  & 	17.254 $\pm$ 0.007 	 & 	16.839 $\pm$  	0.036	 & 	16.514  $\pm$   0.012 \\
B        & 12:04:32.39  & +27:00:47.7     &  18.288 $\pm$ 0.006  & 	17.481 $\pm$ 0.004 	 & 	16.882 $\pm$  	0.028	 & 	16.463  $\pm$   0.010 \\
C        & 12:04:30.35  & +27:00:45.6     &  16.718 $\pm$ 0.002  & 	13.373 $\pm$ 0.003	 & 	16.020 $\pm$  	0.036	 & 	15.716  $\pm$   0.007 \\
D        & 12:05:07.34  & +26:59:24.9     &  18.698 $\pm$ 0.007  & 	17.961 $\pm$ 0.005 	 & 	17.468 $\pm$  	0.015	 & 	17.068  $\pm$   0.022 \\
E        & 12:04:32.88  & +26:58:02.6     &  18.435 $\pm$ 0.007  & 	17.935 $\pm$ 0.004 	 & 	17.532 $\pm$  	0.014	 & 	17.183  $\pm$   0.010 \\
F        & 12:04:44.61  & +26:57:48.9     &  16.008 $\pm$ 0.002  & 	14.998 $\pm$ 0.003 	 & 	14.411 $\pm$  	0.023	 & 	14.021  $\pm$   0.010 \\
G        & 12:05:06.39  & +26:55:46.0     &  18.143 $\pm$ 0.004  & 	17.497 $\pm$ 0.004 	 & 	17.028 $\pm$  	0.017	 & 	16.668  $\pm$   0.008 \\

\hline                                   
\end{tabular}
\label{tab:standard_star_table}      
\end{table*}

\begin{table*}
	\caption{Optical photometry of M12045 in {\it BVRI} bands.}
\centering
\smallskip
\begin{tabular}{c c c c c c c c}
\hline \hline
Date    &   Phase$^\dagger$ &   {\it B}                        &   {\it V}                              &   {\it R}                         &  {\it I}  &  {\it W}                     & Telescope     \\
        &   (Days)          & (mag)                      & (mag)                             & (mag)                       & (mag)     & (mag)                            \\
\hline 
20141028         &   20.87   &  --                      &    --			     &   --		        &  --		       & 14.49$\pm$0.21 & MASTER\\
20141114         &   38.24   &  --                      &    15.04 $\pm$  0.01	     &   14.39 $\pm$ 0.01        &  --		       & -- & ST\\
20141115         &   39.25   &  15.07 $\pm$ 0.02        &    15.12 $\pm$  0.02	     &   14.38 $\pm$ 0.02        &  --		       & -- & ST\\
20141129         &   53.19   &  --                      &    15.28 $\pm$  0.01	     &   --                      &  14.20 $\pm$ 0.01   & -- & ST\\
20141208         &   62.18   &  --    	                &    15.40 $\pm$  0.01	     &   14.83 $\pm$ 0.02        &  14.92 $\pm$ 0.01   & -- & ST\\
20141209         &   63.20   &  17.01 $\pm$ 0.01        &    15.39 $\pm$  0.01	     &   14.80 $\pm$ 0.01        &  14.93 $\pm$ 0.00   & -- & ST\\
20141210         &   64.15   &  17.14 $\pm$ 0.02        &    15.38 $\pm$  0.01	     &   14.84 $\pm$ 0.01        &  14.98 $\pm$ 0.01   & -- & ST\\
20141216         &   70.21   &      -                   &    15.48 $\pm$  0.02	     &   14.93 $\pm$ 0.01        &  15.05 $\pm$ 0.01   & -- & ST\\
20141229         &   83.15   &      -                   &    15.66 $\pm$  0.01	     &   15.15 $\pm$ 0.02        &  15.30 $\pm$ 0.01   & -- & ST\\
20150106         &   91.20   &      -                   &    15.72 $\pm$  0.02	     &   15.24 $\pm$ 0.01        &  15.36 $\pm$ 0.01   & -- & ST\\
20150113         &   97.05   &  --                      &    --			     &   --		        &  --		       & 15.61$\pm$0.12 & MASTER\\
20150114         &   98.04   &  --                      &    --			     &   --		        &  --		       & 15.60$\pm$0.06 & MASTER\\
20150116         &   100.05  &  --                      &    --			     &   --		        &  --		       & 15.61$\pm$0.03 & MASTER\\
20150117         &   101.07  &  --                      &    --			     &   --		        &  --		       & 15.44$\pm$0.06 & MASTER\\
20150122         &   106.89  &  --                      &    --			     &   --		        &  --		       & 15.61$\pm$0.19 & MASTER\\ 
20150124         &   108.68  &  17.16$\pm$ 0.01         &    16.00 $\pm$  0.03	     &   --                      &  15.73 $\pm$ 0.02   & -- & ST\\  
20150130         &   115.10  &  17.23$\pm$ 0.02         &    --               	     &   15.68 $\pm$ 0.01        &  15.80 $\pm$ 0.01   & -- & ST\\ 
20150131         &   116.18  &  17.34$\pm$ 0.01         &    16.14 $\pm$  0.02	     &   15.68 $\pm$ 0.02        &  15.86 $\pm$ 0.01   & -- & ST\\          
20150204         &   120.16  &  17.67$\pm$ 0.03         &    16.09 $\pm$  0.02	     &   15.76 $\pm$ 0.02        &  15.95 $\pm$ 0.02   & -- & ST\\  
20150205         &   121.06  &  17.71$\pm$ 0.01         &    16.19 $\pm$  0.02	     &   15.75 $\pm$ 0.01        &  --                 & -- & ST\\ 
20150216         &   131.82  &  --                      &    --			     &   --		        &  --		       & 15.63$\pm$0.17 & MASTER\\
20150225         &   140.90  &  --                      &    --			     &   --		        &  --		       & 16.10$\pm$0.06 & MASTER\\
20150322         &   165.90  &  17.74$\pm$ 0.02         &    17.19 $\pm$  0.02	     &   16.44 $\pm$ 0.01        &  16.32 $\pm$ 0.02   & -- & ST\\ 
20150323         &   167.08  &  --                      &    17.00 $\pm$  0.02	     &   16.46 $\pm$ 0.01        &  16.74 $\pm$ 0.01   & -- & ST\\
20150327         &   170.80  &  --                      &    --			     &   --		        &  --		       & 16.28$\pm$0.05 & MASTER\\
20150406         &   180.97  &  --                      &    17.26 $\pm$  0.05	     &   16.67 $\pm$ 0.01        &  16.95 $\pm$ 0.01   &  -- & ST\\     
20150407         &   181.94  &  --                      &    17.31 $\pm$  0.02	     &   16.67 $\pm$ 0.01        &  16.90 $\pm$ 0.02   &  -- & ST\\     
20150408         &   182.93  &  --                      &    --                      &   16.70 $\pm$ 0.01        &  16.96 $\pm$ 0.02   &  -- & ST\\ 
20150409         &   184.07  &  --                      &    17.29 $\pm$  0.01       &   --                      &  --                 &  -- & ST\\
20150411         &   186.06  &  --                      &    --                      &   --                      &  16.51 $\pm$ 0.02   &  -- & ST\\
20150418         &   192.93  &      -                   &    17.38 $\pm$  0.02       &   16.89 $\pm$ 0.02        &  17.06 $\pm$ 0.02   &  -- & ST\\  
20150421         &   195.92  &      -                   &    17.44 $\pm$  0.02       &   16.90 $\pm$ 0.02        &  16.95 $\pm$ 0.03   &  -- & ST\\     
20150424         &   198.97  &      -                   &    17.51 $\pm$  0.03       &   16.91 $\pm$ 0.03        &  --                 &  -- & ST\\
20150501         &   205.91  &      -                   &    17.71 $\pm$  0.04       &   17.02 $\pm$ 0.01        &  17.47 $\pm$ 0.03   &  -- & ST\\     
20150502         &   206.92  &      -                   &    17.69 $\pm$  0.02       &   17.12 $\pm$ 0.02        &  17.30 $\pm$ 0.04   &  -- & ST\\
20150503         &   207.96  &      -                   &    17.67 $\pm$  0.02       &   17.05 $\pm$ 0.01        &  17.44 $\pm$ 0.02   &  -- & ST\\
20150605         &   240.65  &       -                  &    --         	     &   --                      &  17.76 $\pm$ 0.05   &  -- & ST\\
20150607         &   242.61  &       -                  &    --        	             &   --                      &  17.67 $\pm$ 0.02   &  -- & ST\\
20150610         &   245.65  &  18.96$\pm$0.01          &    18.33 $\pm$  0.01       &   17.62 $\pm$ 0.01        &  17.82 $\pm$ 0.02   &  -- & HCT\\
20150730         &   295.77  &  --                      &    --			     &   --		        &  --		       & 16.66$\pm$0.06 & MASTER\\

\hline    
\end{tabular}
\newline
	$^\dagger$ Phase has been calculated with respect to $B_{max}$ = 2456938.49 (JD)                                                                                      
\label{tab:photometric_observational_log}                                                        
\end{table*}

\begin{table}
\caption{Log of spectroscopic observations.}
\centering
\smallskip
\begin{tabular}{c c c c c}
\hline \hline
Date          & Phase$^\dagger$          & Grism         & Spectral Range                   \\
              &(Days)                    &               & (\AA)                                     \\
\hline
20141029      &  20.88                & Gr07,Gr08       & 3800-6840,5800-8350         \\
20141111      &  34.94                & Gr07,Gr08       & 3800-6840,5800-8350         \\
20141117      &  40.96                & Gr07,Gr08       & 3800-6840,5800-8350         \\
20141123      &  46.94                & Gr07,Gr08       & 3800-6840,5800-8350         \\
20141201      &  54.90                & Gr07,Gr08 	 & 3800-6840,5800-8350        \\
20141204      &  57.87                & Gr07,Gr08       & 3800-6840,5800-8350         \\
20141219      &  72.84                & Gr07,Gr08       & 3800-6840,5800-8350           \\
20141226      &  79.93                & Gr07,Gr08       & 3800-6840,5800-8350        \\
20150104      &  88.78                & Gr07,Gr08       & 3800-6840,5800-8350         \\
20150106      &  90.80                & Gr07,Gr08       & 3800-6840,5800-8350            \\
20150111      &  95.88                & Gr07,Gr08       & 3800-6840,5800-8350           \\
20150118      &  102.01               & Gr07,Gr08       & 3800-6840,5800-8350           \\
20150125      &  109.76               & Gr07,Gr08       & 3800-6840,5800-8350            \\
20150201      &  116.01               & Gr07,Gr08       & 3800-6840,5800-8350           \\
20150205      &  120.84               & Gr07,Gr08       & 3800-6840,5800-8350            \\
20150207      &  122.83               & Gr07,Gr08       & 3800-6840,5800-8350           \\
20150214      &  129.82               & Gr07,Gr08       & 3800-6840,5800-8350          \\
20150222      &  137.96               & Gr08            & 5800-8350                      \\
20150303      &  146.75               & Gr08            & 5800-8350                     \\
20150325      &  168.81               & Gr07,Gr08       & 3800-6840,5800-8350           \\
20150403      &  177.67               & Gr07,Gr08       & 3800-6840,5800-8350            \\
20150424      &  198.79               & Gr08            & 5800-8350                     \\
20150505      &  209.68               & Gr07,Gr08       & 3800-6840,5800-8350            \\
20150513      &  217.60               & Gr07,Gr08       & 3800-6840,5800-8350            \\
20150522      &  226.75               & Gr07,Gr08       & 3800-6840,5800-8350           \\
20150530      &  234.66               & Gr07            & 3800-6840       	         \\
20150611      &  246.67               & Gr07,Gr08       & 3800-6840		        \\

\hline                                   
\end{tabular}
\newline
	$^\dagger$ Phase has been calculated with respect to $B_{max}$ = 2456938.49 (JD)
\label{tab:spectroscopic_observations}      
\end{table}

\end{document}